\documentclass[a4paper,11pt]{article}

\pdfoutput=1
\usepackage{jheppub_VVK} % for details on the use of the package, please
\usepackage{amsmath}
\usepackage{braket}
\usepackage{graphicx} 

\renewcommand\[{\left[}
\renewcommand\]{\right]}

\def\sst{\scriptscriptstyle}

\def\[{\begin{equation}}
\def\]{\end{equation}}

\title{
Higgs vacuum stability from the dark matter portal
}

\preprint{IPPP/14/22, DCPT/14/44}
\arxivnumber{1403.4953}

\author{Valentin V.~Khoze,}
\author{Christopher McCabe}
\author{and Gunnar Ro}

\affiliation{Institute for Particle Physics Phenomenology, Durham University, \\South Road, Durham, DH1~3LE, U.K.\ }

 \emailAdd{valya.khoze@durham.ac.uk}
 \emailAdd{christopher.mccabe@durham.ac.uk}
  \emailAdd{g.o.i.ro@durham.ac.uk}

\abstract{
We consider classically scale-invariant extensions of the Standard Model ({\it CSI~ESM}) which stabilise the Higgs potential
and have good dark matter candidates.
 In this framework all mass scales, including electroweak and dark matter masses, are generated dynamically and have a common origin. We consider Abelian and non-Abelian hidden sectors portally coupled to the SM with and without a real singlet scalar. We perform a careful analysis of RG running to determine regions in the parameter space where the SM Higgs vacuum is stabilised. After combining this with the LHC Higgs constraints, in models without a singlet, none of the regained parameter space in Abelian ESMs, and only a small section in the non-Abelian ESM survives. However, in all singlet-extended models we find that the Higgs vacuum can be stabilised in all of the parameter space consistent with the LHC constraints. These models naturally contain two dark matter candidates: the real singlet and the dark gauge boson in non-Abelian models. 
 We determine the viable range of parameters in the CSI ESM framework by computing the relic abundance, imposing direct detection constraints and combining with the LHC Higgs constraints. In addition to being instrumental in Higgs stabilisation, we find that the singlet component is required to explain the observed dark matter density.
}

\begin{document}
\maketitle
\flushbottom

%%%%%%%%%%%%%%%%%%%%%%%%%%%%%%%%%%
%%%%%%%%%%%%%%%%%%%%%%%%%%%%%%%%%%
\section{Introduction}\label{sec:intro}
%%%%%%%%%%%%%%%%%%%%%%%%%%%%%%%%%%
%%%%%%%%%%%%%%%%%%%%%%%%%%%%%%%%%%

Following the discovery of the Higgs boson by the ATLAS and CMS experiments 
\cite{ATLAS:2012gk,CMS:2012gu} particle physics has entered a new epoch.
The particle spectrum of the Standard Model is now complete yet nevertheless, we know that the Standard Model cannot be a complete theory
of particle interactions, even if we do not worry about gravity. The more fundamental theory
should be able to address and predict the matter-anti-matter asymmetry of the universe,
the observed dark matter abundance, and it should stabilise the Standard Model Higgs potential.
It should also incorporate neutrino masses and mixings.
In addition it is desirable to have a particle physics implementation of cosmological inflation
and possibly a solution to the strong CP problem. 
Finally there is still a question of the naturalness of the electroweak scale; the Standard Model 
accommodates and provides the description of the Higgs mechanism, but it does not, and 
of course was not meant to, explain the origin of the electroweak scale and why it is so much lighter than the 
UV cut-off scale.

In this paper we concentrate on a particular approach of exploring Beyond the Standard Model (BSM) 
physics, based on the fact that the Standard Model contains a single mass
scale, the negative Higgs mass squared parameter, $-\mu^2_{\sst \rm{SM}}$, in 
the SM Higgs potential,
\begin{equation}
V(H)_{\sst \mathrm{SM}}=-\frac{1}{2}\mu^2_{\sst \mathrm{SM}} H^\dagger H +\lambda_{\sst \mathrm{SM}}( H^\dagger H)^2 \,.
\label{eq:Vsm}
\end{equation}
In the unitary gauge,
$H(x)=\frac{1}{\sqrt{2}}(0, h(x)),$
the vacuum expectation value (vev) $v$ and the mass $m_{h\, \sst \mathrm{SM}}$ of the physical SM Higgs field $h(x)$ are triggered by the
$\mu_{\sst \mathrm{SM}}$ scale,
\begin{equation}
v =\,  \frac{\mu_{\sst \mathrm{SM}}}{(2\lambda_{\sst \mathrm{SM}})^{1/2} }
 \,\,\simeq 246 \,{\rm GeV}\, , \qquad
m_{h\, \sst \mathrm{SM}} = \mu_{\sst \mathrm{SM}} \,\,\simeq 126\, {\rm GeV} \,.
\label{SMfirst}
\end{equation}
If this single mass scale is
generated dynamically in some appropriate extension of the SM, the resulting theory will be manifestly classically scale-invariant. 
Such theories contain no explicit mass-scales ({\it all} masses have to be generated dynamically), 
but allow for non-vanishing beta functions
of their dimensionless coupling constants.  In section {\bf \ref{sec:2}} we employ the seminal mechanism 
of mass-scale generation due to Coleman and Weinberg (CW) \cite{Coleman:1973jx}
and show how the electroweak scale 
emerges in the Standard Model coupled to the CW sector.

Classically Scale-Invariant Extensions of the Standard Model -- {\it CSI ESM} --
amount to a highly predictive model building framework.  
The high degree of predictivity/falsifiability of CSI ESM arises from the fact that one cannot start 
extending or repairing a CSI model by introducing new mass thresholds where new physics
might enter~\cite{Meissner:2006zh, Englert:2013gz}. 
All masses have to be generated dynamically and, at least in the simple models
studied in this paper, 
they are all related to the same dynamical scale, which is not far above the electroweak scale. 
This is consistent with the manifest CSI and as the result protects the electroweak scale itself
by ensuring that there are no heavy mass-scales contributing radiatively to the Higgs mass.
Furthermore,
in the CSI ESM approach one naturally expects the {\it common origin} of all mass scales, i.e. 
the EW scale relevant to the SM, and the scales of new physics. In other words the CSI ESM framework,
if it works, realises the Occam's razor succinctness.

The CSI ESM theory is a minimal extension of the SM which should address 
all the sub-Planckian shortcomings of the SM, such as the 
generation of matter-anti-matter
asymmetry,  dark matter, stabilisation of the SM Higgs potential, neutrino masses, inflation, without introducing scales much higher the electroweak scale.
It was shown recently in Ref.~\cite{Khoze:2013oga} that the CSI
U(1)$_{\sst \mathrm{CW}}\,\times$ \!SM  theory where the
Coleman-Weinberg U(1)$_{\sst \mathrm{CW}}$  sector is re-interpreted as the gauged $\mathrm{B-L}$ U(1) symmetry of the SM, can generate 
the observed value of the matter-anti-matter asymmetry of the Universe without introducing additional mass scales
nor requiring a resonant fine-tuning.
This CSI U(1)$_{\bf \sst  B-L}\,\times$ \!SM theory also generates Majorana masses for the right-handed sterile neutrinos in the range
between 200 MeV and 500 GeV and leads to visible neutrino masses and mixings via the standard sea-saw 
mechanism \cite{Iso:2009ss,Khoze:2013oga}.

It follows that not only the baryonic matter-anti-matter asymmetry, but also the origin of dark matter 
must be related in the CSI ESM to the origin of the electroweak scale and the Higgs vacuum stability.
Papers \cite{Hambye:2013dgv,Carone:2013wla} have shown that in the non-Abelian CSI
SU(2)$_{\sst \mathrm{CW}}\,\times$~\!SM 
theory there is a common origin of the vector dark matter and the electroweak scale.
It was also pointed out in \cite{Khoze:2013uia} that a CSI ESM theory with an additional singlet 
that is coupled non-minimally to gravity, provides a viable particle theory implementation of the
slow-roll inflation. Furthermore, the singlet responsible for inflation also provides an automatic scalar 
dark matter candidate. 

The main motivation of the present paper is to study in detail the link between 
the stability of the electroweak vacuum and the properties of multi-component (vector and scalar) 
dark matter in the context of  CSI ESM theory. Our main phenomenological results are described in 
sections {\bf \ref{sec:Higgs}} and {\bf \ref{sec:DM}}. There, in a model by model basis 
we determine regions on the CSI ESM parameter space where the SM Higgs vacuum 
is stabilised and the extended Higgs sector phenomenology is consistent with the LHC exclusion limits.
We then investigate the dark matter phenomenology, compute the relic abundance and impose constraints
from direct detection for vector and scalar components of dark matter from current and future experiments.

Our discussion and computations in sections {\bf \ref{sec:Higgs}} and {\bf \ref{sec:DM}} are based 
on the CSI EST model-building features and results derived in section {\bf \ref{sec:2}} and
on solving the renormalisation group equations in section {\bf \ref{sec:RGeqs}}.

%%%%%%%%%%%%%%%%%%%%%%%%%%%%%%%%%%
%%%%%%%%%%%%%%%%%%%%%%%%%%%%%%%%%%
\section{CSI ESM building \texorpdfstring{$\&$}{\&} generation of the EW scale}
\label{sec:2}
%%%%%%%%%%%%%%%%%%%%%%%%%%%%%%%%%%
%%%%%%%%%%%%%%%%%%%%%%%%%%%%%%%%%%

In the minimal Standard Model classical scale invariance is broken by the Higgs mass parameter
$\mu^{2}_{\sst \mathrm{SM}}$ in eq.~\eqref{eq:Vsm}. 
Scale invariance is easily restored by reinterpreting this scale
in terms of a vacuum expectation value (vev) of a new scalar $\Phi$, coupled to the SM via the
Higgs portal interaction, $-\,\lambda_{\rm P}|H|^2|\Phi|^{2}.$
Now, as soon as an appropriate non-vanishing value for $\langle \Phi\rangle\ll M_{\sst \mathrm{UV}}$ can emerge 
dynamically, we get  $\mu^2_{\sst \mathrm{SM}} = \lambda_{\rm P}\langle|\Phi|\rangle^2$ in~\eqref{eq:Vsm} 
which
triggers electroweak symmetry breaking.

In order to generate the required vev of $\Phi$ 
we shall follow the approach reviewed in \cite{Englert:2013gz,Khoze:2013uia}
and employ
the seminal mechanism of the mass gap generation due to Coleman and Weinberg~\cite{Coleman:1973jx}. 
In order for the CW approach to be operational, the classical theory should be massless and the scalar field $\Phi$ should be charged under a gauge group $G_{\sst \mathrm{CW}}$. The vev of the CW scalar $\Phi$ appears via the dimensional transmutation
from the running couplings, leading to spontaneous  breaking of $G_{\sst \mathrm{CW}}$ and ultimately to EWSB in the SM. 

The CSI realisations of the Standard Model which we will  concentrate on in this paper 
are thus characterised by the gauge group $G_{\sst \mathrm{CW}}\times {\rm SU(3)}_{\mathrm{c}}\times {\rm SU(2)}_{\mathrm{L}} \times {\rm U(1)}_Y$
where the first factor plays the role of the hidden sector.
The requirement of classical scale invariance implies that
the theory has no input mass scales in its classical Lagrangian; as we already mentioned, all masses have to be generated dynamically via dimensional transmutation.
The basic tree-level scalar potential is
\begin{equation}
V_{\rm cl}(H,\Phi)=\lambda_\phi (\Phi^\dagger \Phi)^2+\lambda_H(H^\dagger H)^2-\lambda_{\rm P}( H^\dagger H)(\Phi^\dagger \Phi)\,.
\label{Vhphi}
\end{equation}
The matter content of the hidden sector gauge group $G_{\sst \mathrm{CW}}$ can vary: 
in the minimal case it consists only of the CW scalar $\Phi$; more generally it can contain additional matter fields, including for example the SM fermions. We will discuss a few representative examples involving Abelian and non-Abelian gauge groups with with a more- and a less-minimal matter content.

The minimal U(1)$_{\sst \mathrm{CW}}$ theory coupled to the SM via the Higgs portal with the scalar potential \eqref{Vhphi} was first considered in
\cite{Hempfling:1996ht}. The
phenomenology of this model was analysed more recently in the context of the LHC, future
colliders and low energy measurements in~\cite{Englert:2013gz}. 
Classical scale invariance is not an exact symmetry of the quantum theory,
but neither is it broken 
by an arbitrary amount. The violation of scale invariance is
controlled by the anomaly in the trace of the energy-momentum tensor, or
equivalently, by the logarithmic running of dimensionless coupling constants and their dimensional 
transmutation scales. In weakly coupled perturbation theory, these are much smaller than the UV cutoff.
Therefore, in order to maintain anomalously broken scale invariance, one should 
select a regularisation scheme that does not introduce explicit powers of the UV cut-off scale~\cite{Bardeen:1995kv}. 
In the present paper we use dimensional regularisation with the $\overline{\rm MS}$ scheme.  
In dimensional regularisation, and in theories like ours that
  contain no explicit mass scales at the outset, no large
  corrections to mass terms can appear. In this regularisation, which preserves classical
  scale invariance, the CSI ESM theory is not fine-tuned in the technical sense 
  \cite{Khoze:2013uia,Englert:2013gz}.

Other related studies of CSI ESModels can be found in~\cite{Chang:2007ki,Foot:2007ay,Foot:2007iy,Holthausen:2009uc,Iso:2012jn,Bian:2013wna,Guo:2014bha}.
We would also like to briefly comment on two scale-invariance-driven approaches which 
are different from ours. The authors of Refs.~\cite{Meissner:2006zh,Hambye:2007vf,Gabrielli:2013hma,Holthausen:2013ota,Hill:2014mqa} envision 
CSI models with dimensional transmutation, which are not based on the CW gauge-sector-extension of the SM, but
rather appeal to an extended matter content within the SM, or to a strongly coupled hidden sector.
One can also consider model building based on the approach with an exact quantum scale invariance of the UV theory,
as discussed recently in \cite{Tavares:2013dga} and \cite{Abel:2013mya}. It is important to keep in mind that
classical scale invariance of the effective theory below the Planck scale does not necessarily assume or is 
directly related to the hypothesised conformal invariance of the UV embedding of the SM.

%%%%%%%%%%%%%%%%%%%%%%%%%%%%%%%%%%
\subsection{CSI \texorpdfstring{U(1)$_{\bf \sst \mathrm{CW}}\,\times$ \!SM}{U(1)CWSM}}
\label{sec2:U1CW}
%%%%%%%%%%%%%%%%%%%%%%%%%%%%%%%%%%

This is the minimal classically scale-invariant extension of the SM. The SM Higgs doublet $H$
is coupled via the Higgs-portal interactions to the complex scalar 
\[
\Phi \,=\, \frac{1}{\sqrt{2}}(\phi + i \phi_2)\,,
\]
where $\Phi$ is a Standard Model singlet, but
charged under the U(1)-Coleman-Weinberg gauge group. The hidden sector consists of this U(1) with $\Phi$
plus nothing else. In the unitary gauge one is left with two real scalars,
\[
H=\frac{1}{\sqrt{2}}(0,h)\, , \quad 
\Phi=\frac{1}{\sqrt{2}}\phi\,,
\]
and the tree-level scalar potential \eqref{Vhphi} reads
\begin{equation}
V_0(h,\phi)=\frac{\lambda_{\phi}^{(0)}}{4}\phi^4+
\frac{\lambda_H^{(0)}}{4}h^4
-\frac{\lambda_{\rm P}^{(0)}}{4} h^2 \phi^2\,,
\label{V0hphi}
\end{equation}
where the superscripts indicate that the corresponding coupling constants are  the tree-level quantities.

We now proceed to include radiative corrections to the classically scale-invariant potential above.
Our primary goal in this section is to show how quantum effects generate the non-trivial vacuum with non-vanishing
vevs
$\langle \phi\rangle$ and $v=\langle h \rangle$, to derive the matching condition 
between coupling constants in the vacuum and to compute the scalar mass eigenstates, $m_{h_1}^2$ and $m_{h_2}^2$
of the mixed scalar fields $h$ and $\phi$. We then determine the SM Higgs self-coupling $\lambda_{\sst \mathrm{SM}}$ in terms of
$\lambda_{H}$ and other parameters of the model. The fact that $\lambda_{\sst \mathrm{SM}}$ is not identified with $\lambda_{H}$
will be of importance later when we discuss the stability of the SM Higgs potential in our model(s).

For most of this section we will follow closely the analysis of Ref.~\cite{Englert:2013gz}, but with a special emphasis
on two aspects of the derivation. First, is that the effective potential and the running couplings need to be
computed in the $\overline{\rm MS}$ scheme, which is the scheme we will also use later on for writing down and solving the RG equations. 

Following the approach outlined in \cite{Englert:2013gz} one can simplify the derivation considerably by 
first concentrating primarily on the CW sector and singling out the 1-loop contributions $\propto e_{\sst \mathrm{CW}}^4$
arising from the hidden U(1) gauge field.\footnote{Radiative corrections due to the CW scalar self-coupling 
$\propto \lambda_{\phi}^2$ will be sub-leading in this approach cf.~eq.~\eqref{eq:cwmsbar} below.
}
Perturbative corrections arising from the SM sector will then be added later. 
Effective potentials and  running couplings in this paper will always be computed in the $\overline{\rm MS}$ scheme. 
In this scheme the 1-loop effective potential for $\phi$ 
reads, cf.~\cite{Q},
\[
V_1(\phi;\mu)=\frac{\lambda^{(0)}_\phi}{4}\,\phi^4 \,+\,\frac{3}{64\pi^2}\,e^4_{\sst \mathrm{CW}}(\mu)\,\phi^4\left(\,\log\frac{
e^2_{\sst \mathrm{CW}}(\mu)\,\phi^2}{\mu^2}-\frac{5}{6}\right)
\,,
\label{V1one}
\]
which depends on the RG scale $\mu$ that appears both in the logarithm and 
also in the 1-loop running CW gauge coupling constant $e_{\sst \mathrm{CW}}(\mu)$.
The running (or renormalised) self-coupling $\lambda_{\phi} $ at the RG scale $\mu$ is defined via
\[
\lambda_{\phi} (\mu) \,=\, \frac{1}{3!}\,\left(\frac{\partial^4 V_1 (\phi;\mu)}{\partial \phi^4}\right)_{\phi=\mu}
 \,=\, 
 \lambda_{\phi}^{(0)} \,+\,
  \frac{10 e_{\sst \mathrm{CW}}(\mu)^4 +3 e_{\sst \mathrm{CW}}(\mu)^4 \log \left(e_{\sst \mathrm{CW}}(\mu)^2\right)}{16 \pi ^2}\,.
  \label{lphiR1}
\]

We can now express the effective potential in terms of this renormalised coupling constant
by substituting 
$\lambda_\phi^{(0)} \,=\, 
 \lambda_{\phi} \,-\, (10 e_{\sst \mathrm{CW}}^4+3 e_{\sst \mathrm{CW}}^4 \log e_{\sst \mathrm{CW}}^2)/{(16 \pi^2)}$
 into eq.~\eqref{V1one}, obtaining
\[
V_1 (\phi;\mu)\,=\, 
\frac{\lambda_\phi(\mu) \phi^4}{4}+ \frac{3e_{\sst \mathrm{CW}}(\mu)^4}{64 \pi ^2}  \phi^4
\left(\log \left(\frac{\phi^2}{\mu^2}\right)-\frac{25}{6}\right) \,.
\label{V1R}
\]
The vacuum of the effective potential above occurs at $\langle \phi \rangle \neq 0.$ 
Minimising the potential \eqref{V1R} with respect to $\phi$ at $\mu=\langle \phi \rangle$ gives the 
characteristic Coleman-Weinberg-type $\lambda_\phi \propto e_{\sst \mathrm{CW}}^4$ relation between 
the scalar and the gauge couplings,
\[
\lambda_\phi \,=\, \frac{11}{16\pi^2} \,e_{\sst \mathrm{CW}}^4 \qquad {\rm at} \quad \mu=\langle \phi\rangle\,.
\label{eq:cwmsbar}
\]
It is pleasing to note that this matching relation between the couplings takes exactly the same form as the one
obtained in the CW paper in the cut-off scheme -- i.e. accounting for the 3! mismatch in the definition of
the coupling in \cite{Coleman:1973jx} we have $\lambda_\phi=\frac{1}{3!} \lambda$ where $\lambda$ is the
coupling appearing in \cite{Coleman:1973jx}, $\lambda=\frac{33}{8\pi^2}e^4$.

Shifting the CW scalar by its vev $\phi \to \langle \phi \rangle + \phi$ and expanding the effective potential
in \eqref{V1R}, we
find the mass of $\phi$,
\[m_\phi^2\,=\,
\frac{3e_{\sst \mathrm{CW}}^4}{8\pi^2}\langle \phi \rangle^2\,,
\label{eq:mphiZ}
\]
and the mass of the $Z'$ U(1) vector boson,
\[
M_{Z'}^2 = e_{\sst \mathrm{CW}}^2 \langle \phi \rangle^2\,
\quad \gg \quad 
m_\phi^2 =\frac{3e_{\sst \mathrm{CW}}^4}{8\pi^2}\langle \phi \rangle^2\,
\,.
\label{masses2}
\]
The $\overline{\rm MS}$ expressions above are once again identical to those derived in the cut-off scheme 
in~\cite{Coleman:1973jx,Englert:2013gz}.

We now turn to the SM part of the scalar potential \eqref{V0hphi}, specifically
\begin{equation}
V_0(h)=
\frac{\lambda_H^{(0)}}{4}h^4
-\frac{\lambda_{\rm P}\langle \phi\rangle^2}{4} h^2 \,,
\label{VSM0hphi}
\end{equation}
where we have dropped the $(0)$ superscript for the portal coupling, as it will turn out that
$\lambda_{\rm P}$ does not run much.
The SM scale $\mu^2_{\sst \mathrm{SM}}$ is generated by the CW vev in the second term,
\[
\mu^2_{\sst \mathrm{SM}}=\lambda_{\rm P} \langle \phi\rangle^2\,,
\]
 and this triggers in turn the 
appearance of the Higgs vev $v$ as in the first equation in \eqref{SMfirst}.

The presence of the portal coupling in the potential \eqref{VSM0hphi} (or more generally \eqref{V0hphi}) provides a correction
 to the CW matching condition 
\eqref{eq:cwmsbar} and the CW mass \eqref{eq:mphiZ}.
By including the last term on the {\it r.h.s} of
\eqref{V0hphi} to the effective potential in \eqref{V1one} and \eqref{V1R}, we find the $\lambda_{\rm P}$-induced correction
to the equations \eqref{eq:cwmsbar}-\eqref{eq:mphiZ} which now read 
\begin{align}
\lambda_\phi &= \frac{11}{16\pi^2} \,e_{\sst \mathrm{CW}}^4 
+\lambda_{\rm P}\frac{v^2}{2\langle\phi\rangle^2}
\qquad {\rm at} \quad \mu=\langle \phi\rangle
\label{eq:cwmsbar-P}
\\
m_\phi^2 &=
\frac{3e_{\sst \mathrm{CW}}^4}{8\pi^2}\langle \phi \rangle^2
+\lambda_{\rm P} v^2
\label{eq:mphiZ-P}
\end{align}
in full agreement with the results of \cite{Englert:2013gz}. In this paper, we consider small values of $\lambda_{\rm P}$ so that these corrections are negligible, since
$\lambda_{\rm P} v^2/(2 \langle \phi \rangle^2) \sim \lambda_{\rm P}^2/(4 \lambda_{H}) \ll 1.$

Our next task is to compute the Higgs mass including the SM radiative corrections.
To proceed we perform the usual shift, $h(x) \to v+h(x)$,
and represent the SM scalar potential \eqref{VSM0hphi} as follows,
\begin{equation}
V(h)=
\frac{\lambda_H^{(0)}}{4}(v+h)^4
-\frac{\mu^2_{\sst \mathrm{SM}}}{4} (v+h)^2 
+\frac{1}{2} \Delta m^{2}_{h,\rm{SM}} \, h^2
\,,
\label{VSMH}
\end{equation}
where for overall consistency we have also included one-loop corrections
to the Higgs mass arising in the Standard Model,
\begin{equation}
  \Delta m^{2}_{h,\rm{SM}}=\frac{1}{16\pi^2}\frac{1}{v^2}\left(6m^{4}_{W}+3m^{4}_{Z}+m^{4}_{h}-24m^{4}_{t}\right)
  \approx -2200\, {\rm GeV}^{2}\,.
\end{equation}
These corrections are dominated by the
top-quark loop and are therefore negative. The appearance of $v^2$ in the denominator of
$ \Delta m^{2}_{h,\rm{SM}}$ is slightly misleading, and it is better to recast it as,
\begin{equation}
  \Delta m^{2}_{h,\rm{SM}}= 2\Delta \lambda_H \, v^2 \, , \quad
  {\rm where}\quad
  \Delta \lambda_H \simeq - 0.018\,.
\end{equation}
The vev $v$ is determined from \eqref{VSMH} by minimisation and setting $h(x)=0$, and thus the last
term in \eqref{VSMH} does not affect the value of $v$, however it does contribute to the one-loop 
corrected value of the Higgs mass. We have,
\[
v^2=\frac{\lambda_{\rm P}}{2\lambda^{(0)}_H}\, \langle \phi \rangle^2
\, , \qquad 
m_h^2 = 2\lambda_H\, v^2 
\, , \qquad \lambda_H = \lambda^{(0)}_H + \Delta \lambda_H \simeq \lambda^{(0)}_H - 0.018 
\,,
\label{masses1}
\]
where $\lambda_H$ is the one-loop corrected value of the self-coupling.

The two
scalars, $h$ and $\phi$, both have vevs and hence mix via the mass matrix, 
\begin{equation}
\label{Mmixing}
  M^{2}=\left(
\begin{array}{cc}
2\lambda_H\, v^2   &  - \sqrt{2\lambda_{\rm P} \lambda^{(0)}_{H}} v^2
\\
% & \\
 - \sqrt{2\lambda_{\rm P} \lambda^{(0)}_{H}} v^2  &  m^{2}_{\phi} 
\end{array}\right)\,,
\end{equation}
where $m^{2}_{\phi}$ is given in 
\eqref{eq:mphiZ-P} (and already includes the $\lambda_{\rm P}$ correction).\footnote{The mass mixing matrix \eqref{Mmixing}
 is equivalent to the mass matrix derived in \cite{Englert:2013gz} which was of the form:
 %\begin{equation}
$
M^{2}=\left(
\begin{array}{cc}
  m^{2}_{h,0}+\Delta m^{2}_{h,\rm SM} & -\kappa \,m^{2}_{h,0} \\
  -\kappa\, m^{2}_{h,0}  &  m^{2}_{\phi,0} +\kappa^{2} m^{2}_{h,0}
\end{array}\right)$ 
in terms of 
$m^{2}_{h,0}= \, {2\lambda^{(0)}_H}\, v^2$ and $m_{\phi,0}^2 =
3e_{\sst \mathrm{CW}}^4\langle \phi \rangle^2/(8\pi^2)$,
with
$\kappa=\sqrt{\lambda_{\rm P}/(2\lambda^{(0)}_{H})}.$
%\end{equation}
}
The mass eigenstates are the two Higgs fields, $h_1$ and $h_2$ with the mass eigenvalues,
\[
m_{h_1,h_2}^2=\frac{1}{2} \left(2 \lambda _H v^2 +m_{\phi }^2 
\pm \sqrt{\left(2 \lambda _H v^2 -m_{\phi }^2\right)^2  +8\lambda_{\rm P} \lambda _{H}^{(0)} v^4}\right)\,.
\label{mh1h2}
\]
It is easy to see that in the limit where the portal coupling $\lambda_{\rm P}$ is set to zero, 
the mixing between the
two scalars $h$ and $\phi$ disappears resulting in $m_h^2$ and $m_\phi^2$ mass eigenvalues, as one would expect.
However, for non-vanishing $\lambda_{\rm P}$, the mass eigenstates $h_1$ and $h_2$ are given by
\[ \left( \begin{array}{cc}
h_1\\
h_2
 \end{array} \right)=\left(\begin{array}{cc}
\cos\, \theta & -\sin \,\theta\\
\sin\, \theta &\, \,\,\,\cos \,\theta
 \end{array} \right) \left( \begin{array}{cc}
h\\
\phi
 \end{array} \right)\]
 with a nontrivial mixing angle $\theta$.
 Which of these two mass eigenstates should be identified with the SM Higgs 
$m^2_{h\, \sst \mathrm{SM}} = \,\simeq (126\, {\rm GeV} )^2 $ of eq.~\eqref{SMfirst}? 

The answer is obvious, the SM Higgs is the
eigenstate $h_1$ which is `mostly' the $h$ scalar (i.e. $\cos \theta \times$the scalar coupled to the SM electroweak sector) for small values of the mixing angle,
\[
h_{\sst \mathrm{SM}} \,:=\, h_1 \,=\, h\, \cos\, \theta \,-\, \phi\, \sin\, \theta\,  , \qquad 
m_{h_1}  = 125.66\, {\rm GeV} \,.
\label{SMh1-ident}
\]
The SM Higgs self-coupling constant $\lambda_{\sst \mathrm{SM}}$ appearing in the SM Higgs potential \eqref{eq:Vsm}
can be inferred from
$ m_{h_1}^2 = 2 \lambda_{\sst \mathrm{SM}} v^2$,
but it is not the relevant or primary parameter in our model ($\lambda_H$ is).

In our computations for the RG evolution of couplings and the analysis of Higgs potential stabilisation carried out 
in this paper, we solve the initial condition \eqref{SMh1-ident} for the eigenvalue problem of \eqref{Mmixing}
numerically without making analytical approximations. However, we show some simple analytic expressions to illuminate our approach.

In the approximation where $(8\lambda_{\rm P} \lambda^{(0)}_H v^4 )/(2\lambda_H v^2-m_\phi^2)^2$ is small
we can expand the square root in \eqref{mh1h2} and obtain:
\begin{align}
m^2_{h_1} \,=\, m^2_{+} &=
2v^2\lambda_H \left(1+\frac{\lambda_{\rm P}(\lambda_H^{(0)}/\lambda_H)\,v^2}{2\lambda_H v^2-m_\phi^2}\right)\,,
\quad{\rm for}~~2\lambda_H v^2 >m_\phi^2\,,
\label{m2+} \\
m^2_{h_1} \,=\, m^2_{-} &=
2v^2\lambda_H \left(1-\frac{\lambda_{\rm P}(\lambda_H^{(0)}/\lambda_H)\,v^2}{m_\phi^2-2\lambda_H v^2}\right)\,,
\quad{\rm for}~~m_\phi^2>2\lambda_H v^2 \,.
\label{m2-}
\end{align}
Note that our requirement of assigning the SM Higgs mass value of 126 GeV to the `mostly $h$ state'
selects two different roots of \eqref{mh1h2} in the equations above, depending on whether the $h$ state
or the $\phi$ state is lighter. As the result, there is a `discontinuity of the SM Higgs identification' with
$m^2_{h_1} > 2v^2\lambda_H$ in the first equation, while
$m^2_{h_1} < 2v^2\lambda_H$ in the second equation. 
Similarly, the value of $\lambda_{H}$ is smaller or greater than the perceived value of $\lambda_{\sst \mathrm{SM}}$ in the SM,
in particular,
\begin{equation}
\lambda_{\sst \mathrm{SM}}=
\lambda_H \left(1-\frac{\lambda_{\rm P}(\lambda_H^{(0)}/\lambda_H)\,v^2}{m_\phi^2-2\lambda_H v^2}\right)\,,
\quad{\rm for}~~m_\phi^2>2\lambda_H v^2 \,.
\label{lSM-}
\end{equation}
One concludes that in the case of the CW scalar being heavier than the SM Higgs, it should be easier to stabilise
the SM Higgs potential, since the initial value of $\lambda_H$ here is larger than the initial value of the 
$\lambda_{\sst \mathrm{SM}}$ coupling and as such, it should be useful in preventing $\lambda_H$ from going negative
 at high values of
the RG scale.\footnote{This point has been noted earlier in the literature in
 \cite{Lebedev:2012zw,EliasMiro:2012ay}, \cite{Hambye:2013dgv}
in the context of assisting the stabilisation of the
SM Higgs by integrating out a heavy scalar. In our case the second scalar does not have to be integrated out. In fact, 
the
required stabilising effect arises when the second scalar is not much heavier than the SM Higgs, which manifests itself in 
keeping the denominator in \eqref{lSM-} not much greater than the square of the EW scale.}

On a more technical note, in our computations we also take into account the fact that
the requirement of stability of the Higgs potential at high scales 
goes beyond the simple condition 
$\lambda_{H} (\mu)> 0$ at all values of $\mu$, but should be supplemented by the slightly stronger requirement emerging
from the tree-level stability of the potential \eqref{V0hphi}, which requires that $\lambda_{H} > \lambda_{\rm P}^2/(4 \lambda_\phi).$

In the following sections {\bf \ref{sec2:U1BL}-\ref{sec2:U1BLsc}}, we extend the construction above
to models with  more general hidden sectors. First of all, the G$_{\sst \mathrm{CW}}$ Coleman-Weinberg sector can 
be extended so that SM fermions are charged under G$_{\sst \mathrm{CW}}$, and, secondly, G$_{\sst \mathrm{CW}}$ can also be non-Abelian. In addition, these
CSI ESM models can include a gauge singlet with portal couplings to the Higgs and the CW scalar field.
In sections 
{\bf \ref{sec:Higgs}} and {\bf \ref{sec:DM}}
we will explain how the combination of constraints arising from the Higgs vacuum stability, collider exclusions, and dark matter 
searches and phenomenology will apply to and discriminate between these varieties of CSI SM extensions.

%%%%%%%%%%%%%%%%%%%%%%%%%%%%%%%%%%
\subsection{CSI \texorpdfstring{U(1)$_{\bf \bf \sst  B-L}\,\times$ \!SM}{U(1) B-L SM}}
\label{sec2:U1BL}
%%%%%%%%%%%%%%%%%%%%%%%%%%%%%%%%%%

The $\mathrm{B-L}$ theory was originally introduced in \cite{Mohapatra:1980qe},
and in the context of the CW classically scale-invariant extension of the SM this theory
was recently studied in \cite{Iso:2012jn} and by the two of the present authors in
\cite{Khoze:2013oga}. In the latter reference it was shown that this model can 
explain the matter-antimatter asymmetry of the universe by adopting
the `Leptogenesis due to neutrino oscillations' mechanism of \cite{Akhmedov:1998qx}
in a way which is consistent with the CSI requirement that there are no large mass scales
present in the theory.

The U(1)$_{\bf \bf \sst  B-L}\times$ SM theory is a particularly appealing CSI ESM realisation, since the
gauge anomaly of  U(1)$_{\bf \sst  B-L}$ cancellation requires that the matter content of the model
automatically includes
three generations of right-handed Majorana neutrinos.
All SM matter fields are charged under the U(1)$_{\bf \sst  B-L}$ gauge group with charges equal to their Baryon minus 
Lepton number. In addition, 
the CW field $\phi$ carries the $\mathrm{B-L}$ charge 2 and its vev generates the
Majorana neutrino masses and the mass of the U(1)$_{\bf \sst  B-L}$ $Z'$ boson.
The standard see-saw mechanism generates masses of visible neutrinos and also leads to neutrino oscillations.

The scalar field content of the model is the same as before,
with $H$ being the complex doublet and $\Phi \,=\, \frac{1}{\sqrt{2}}(\phi + i \phi_2)$, the complex singlet under the SM.
The tree-level scalar potential is given by \eqref{Vhphi} which in the unitary gauge takes the form \eqref{V0hphi}.
Our earlier discussion of the mass gap generation in the CW sector, the EWSB 
and the mass spectrum structure,
 proceeds precisely as in the previous sections, with the substitution
$e_{\sst \mathrm{CW}} \to \,2\,e_{\bf \sst  B-L}$. The one-loop corrected potential \eqref{V1R}
becomes:
\[
V_1(\phi)=\frac{\lambda_\phi(\mu)}{4}\phi^4 +\frac{3}{64\pi^2}(2e_{\bf \sst  B-L}(\mu))^4\phi^4\left(\log\frac{\phi^2}{\mu^2}-\frac{25}{6}\right)
-\frac{\lambda_{\rm P}(\mu)}{4} h^2 \phi^2\,.
\]
Minimising it at $\mu=\left<\phi\right>$ gives the matching condition for the couplings and the expansion around the
vacuum at $\left<\phi\right>$  determines the mass of the CW scalar field (cf.~\eqref{eq:cwmsbar-P}-\eqref{eq:mphiZ-P}),
\begin{align}
\lambda_\phi &= \frac{11}{\pi^2} \,e_{\bf \sst  B-L}^4 
+\lambda_{\rm P}\frac{v^2}{2\langle\phi\rangle^2}
\qquad {\rm at} \quad \mu=\langle \phi\rangle
\label{eq:cwmsbar-PBL}
\\
m_\phi^2 &=
\frac{6e_{\bf \sst  B-L}^4}{\pi^2}\langle \phi \rangle^2
+\lambda_{\rm P} v^2
\label{eq:mphiZ-PBL}
\end{align}
in agreement with \cite{Khoze:2013oga}. The expressions for the Higgs field vev, $v$, and the Higgs mass, $m_h$, 
are unchanged and given by  \eqref{masses1}. The mass mixing matrix is the same as in \eqref{Mmixing}
with $m_\phi^2$ given by \eqref{eq:mphiZ-PBL}.

%%%%%%%%%%%%%%%%%%%%%%%%%%%%%%%%%%
\subsection{CSI \texorpdfstring{SU(2)$_{\bf \sst \mathrm{CW}}\,\times$ \!SM}{SU(2) CW SM}}
\label{sec2:SU2CW}
%%%%%%%%%%%%%%%%%%%%%%%%%%%%%%%%%%

One can also use a
non-Abelian extension of the SM in the CSI ESM general framework.
In this section we concentrate on the simple case where the
CW group is SU$(2)$ and for simplicity we assume that there are no additional matter fields (apart from the CW scalar~$\Phi$) charged under this hidden sector gauge group. This model was previously considered in~\cite{Hambye:2013dgv} and subsequently in
\cite{Carone:2013wla}. The novel feature of this model is the presence of the vector dark matter candidate -- the 
SU$(2)$ Coleman-Weinberg gauge fields \cite{Hambye:2013dgv}.

The classical scalar potential is the same as before,
\[V_{\rm cl}
(H,\Phi)=\lambda_\phi (\Phi^\dagger \Phi)^2+\lambda_H( H^\dagger H)^2-\lambda_{\rm P}(H^\dagger H)(\Phi^\dagger \Phi)\,,
\label{VSU2}
\]
where $\Phi$ as well as the Higgs field $H$ are the complex doublets of the SU(2)$_{\sst \mathrm{CW}}$ and the
SU(2)$_{\sst \mathrm{L}}$ respectively. 
In the unitary gauge for both of the SU(2) factors we have,
\[
H=\frac{1}{\sqrt{2}}(0,h),\Phi=\frac{1}{\sqrt{2}}(0,\phi)\,.
\]
The analogue of the one-loop corrected scalar potential \eqref{V1R} now 
becomes:
\[
V_1(\phi)=\frac{\lambda_\phi(\mu)}{4}\phi^4 +\frac{9 }{1024\,\pi^2}\, g^4_{\sst \mathrm{CW}}(\mu)
\,\phi^4\left(\log\frac{\phi^2}{\mu^2}-\frac{25}{6}\right)
-\frac{\lambda_{\rm P}(\mu)}{4} h^2 \phi^2\,,
\]
where $g_{\sst \mathrm{CW}}$ is the coupling of the SU(2) CW gauge sector.
Minimising at $\mu=\left<\phi\right>$ gives:
\begin{align}
\lambda_\phi &= \frac{33}{256\,\pi^2} \,g_{\sst \mathrm{CW}}^4 
+\lambda_{\rm P}\frac{v^2}{2\langle\phi\rangle^2}
\qquad {\rm at} \quad \mu=\langle \phi\rangle
\label{eq:cwmsbar-PSU2}
\\
m_\phi^2 &=
\frac{9 }{128\,\pi^2}\,g_{\sst \mathrm{CW}}^4\, \langle \phi \rangle^2
+\lambda_{\rm P} v^2\,.
\label{eq:mphiZ-PSU2}
\end{align}

%%%%%%%%%%%%%%%%%%%%%%%%%%%%%%%%%%
\subsection{CSI ESM \texorpdfstring{$\oplus$}{plus} singlet}
\label{sec2:U1BLsc}
%%%%%%%%%%%%%%%%%%%%%%%%%%%%%%%%%%

All Abelian and non-Abelian CSI extensions of the SM introduced above
can be easily extended further by adding a singlet degree of freedom,  a one-component real scalar field $s(x)$. 
Such extensions by a real scalar were recently shown in Ref.~\cite{Khoze:2013uia} to be 
instrumental in generating the slow-roll potential for cosmological inflation 
when the scalar $s(x)$ is coupled non-minimally to gravity.
The two additional features of models with the singlet, which are particularly important for the purposes of this paper, 
are that (1) the singlet portal coupling to the Higgs will provide an additional (and powerful) potential 
for the Higgs stabilisation, and (2) that
the singlet $s(x)$ is also a natural candidate for scalar dark matter.

The gauge singlet $s$ field is coupled to the ESM models of sections {\bf \ref{sec2:U1CW}-\ref{sec2:SU2CW}}
 via  the scalar portal interactions with the Higgs and the CW field $\Phi$,
  \begin{equation}
    \label{potentialcoupled3}
    V_{\rm cl}(H,\phi,s)\,=\,  \frac{\lambda_{Hs}}{2}H^\dagger H s^2
    \,+\, \frac{\lambda_{\phi s}}{2} \Phi^\dagger \Phi s^2
    \,+\, \frac{\lambda_{s}}{4}  s^4 
     \,+\,   V_{\rm cl}(H,\Phi)\,.
  \end{equation}
Equations~\eqref{Vhphi},~\eqref{potentialcoupled3} describe the general renormalisable gauge-invariant scalar potential 
for the three classically massless scalars as required by classical scale invariance. 
The coupling constants in the potential \eqref{potentialcoupled3} are taken to be all positive, thus the potential is stable and
the positivity of $\lambda_{Hs}$ and $\lambda_{\phi s}$ ensure that no vev is generated for the singlet $s(x)$. Instead the 
CW vev $\langle \phi \rangle$ generates the mass term for the singlet,
\begin{equation}
    \label{ms2}
  m_s^2\,=\,  \frac{\lambda_{Hs}}{2}\,v^2 
    \,+\, \frac{\lambda_{\phi s}}{2} \, |\langle \phi \rangle|^{2}
    \,,
  \end{equation}
in the vacuum $s=0$, $\phi=\langle \phi \rangle$, $H=\frac{v}{\sqrt{2}}=\,\sqrt{ \frac{\lambda_{\rm P}}{\lambda_{\rm H}}}\, |\langle \phi \rangle |$.

%%%%%%%%%%%%%%%%%%%%%%%%%%%%%%%%%%%%%%%%%%%%%%%%%%%%
%%%%%%%%%%%%%%%%%%%%%%%%%%%%%%%%%%%%%%%%%%%%%%%%%%%%
\section{RG Evolution}
\label{sec:RGeqs}
%%%%%%%%%%%%%%%%%%%%%%%%%%%%%%%%%%%%%%%%%%%%%%%%%%%%
%%%%%%%%%%%%%%%%%%%%%%%%%%%%%%%%%%%%%%%%%%%%%%%%%%%%
In this section our aim is to put together a tool kit which will be necessary to determine regions
of the parameter spaces of CSI ESModels where the Higgs vacuum is stable. To do this we first need 
to specify the RG equations 
for all CSI ESM theories of interest, with and without the additional singlet.
We will also fix the initial conditions for the RG evolution.

Following this more technical build up in the present section, the Higgs vacuum stability and collider constraints on the Higgs-sector phenomenology will be analysed 
in section~{\bf \ref{sec:Higgs}}.

%%%%%%%%%%%%%%%%%%%%%%%%%%%%%%%%%%%%%%%%%%%%%%%%%%%%
\subsection{Standard Model \texorpdfstring{$\times$ U(1)${_{\bf \sst  CW}}$}{xU(1)CW}}
%%%%%%%%%%%%%%%%%%%%%%%%%%%%%%%%%%%%%%%%%%%%%%%%%%%%

This is the simplest scale-invariant extension of the SM. The hidden sector is an Abelian U(1) which couples
only to the CW scalar (of charge 1) and no other matter fields. 
We now proceed to write down the renormalisation group equations for this model.

The scalar couplings $\lambda_H$, $\lambda_\phi$ and $\lambda_{\rm P}$ are governed by:
\begin{eqnarray}
(4\pi)^2 \frac{d \lambda_H}{d \log \mu}&=&-6 y_t^4+24\lambda_H^2+ \lambda_{\rm P}^2 +
\lambda_H \left(12y_t^2-\frac{9}{5}g_1^2-9g_2^2 -3g_{\rm mix}^2\right) 
 \nonumber\\ 
 &&+\frac{27}{200}g_1^4+\frac{9}{20}g_2^2g_1^2
+\frac{9}{8}g_2^4
+\frac{3}{4}g_2^2g_{\rm mix}^2
 +\frac{9}{20}g_1^2g_{\rm mix}^2+\frac{3}{8}g_{\rm mix}^4
\label{lHU1}\\ 
\nonumber \\
(4\pi)^2 \frac{d \lambda_\phi}{d \log \mu} &=& 20\lambda_\phi^2 +2\lambda_{\rm P}^2
-12\lambda_\phi \, e_{\sst \mathrm{CW}}^2 +6 e_{\sst \mathrm{CW}}^4\\ 
(4\pi)^2 \frac{d \lambda_{\rm P}}{d \log \mu}&=&\lambda_{\rm P}\left(6y_t^2
+12\lambda_H+8\lambda_\phi -4\lambda_{\rm P}
-6e_{\sst \mathrm{CW}}^2
 -\frac{9}{10}g_1^2  -\frac{9}{2}g_2^2
-\frac{3}{2}g_{\rm mix}^2\right)
-3g_{\rm mix}^2e_{\sst \mathrm{CW}}^2
\nonumber\\
\end{eqnarray}
The RG equation for the top Yukawa coupling $y_t$ is,
\begin{eqnarray}
(4\pi)^2 \frac{d y_t}{d \log \mu}=y_t\left(\frac{9}{2}y_t^2 -\frac{17}{20}g_1^2-\frac{9}{4}g_2^2-8g_3^2
-\frac{17}{12}g_{\rm mix}^2
\right)\,.
\end{eqnarray}
Finally, $e_{\sst \mathrm{CW}}$,  $g_{\rm mix}$ and $g_{i}$  denote the gauge couplings of the U(1)$_{\sst \mathrm{CW}} \times \mathrm{SM}$,
which obey,
\begin{eqnarray}
&& (4\pi)^2 \frac{d e_{\sst \mathrm{CW}}}{d \log \mu}=\frac{1}{3}e_{\sst \mathrm{CW}}^3+\frac{41}{6}e_{\sst \mathrm{CW}}g_{\rm mix}^2
\\
&&(4\pi)^2 \frac{d g_{\rm mix}}{d \log \mu}=\frac{41}{6}g_{\rm mix}\left(g_{\rm mix}^2+2g_1^2\right)+\frac{1}{3}e_{\sst \mathrm{CW}}^2g_{\rm mix}\\
&&(4\pi)^2 \frac{d g_3}{d \log \mu}=-7g_3^3 \,,\quad
(4\pi)^2 \frac{d g_2}{d \log \mu}=-\frac{19}{6}g_2^3  \,,\quad
(4\pi)^2 \frac{d g_1}{d \log \mu}=\frac{41}{10}g_1^3\,.
\label{gaugeU1}
\end{eqnarray}
A characteristic feature of the Abelian ESM theory is $g_{\rm mix}$, the kinetic mixing  of the two Abelian factors,
 U(1)$_{\sst \mathrm{CW}} \times {\rm U(1)}_Y$.
For a generic matter
field $\varphi$ transforming under both U(1)'s with the charges $Q^{\sst \mathrm{CW}}$ and $Q^{Y}$, the kinetic mixing is defined 
as the coupling constant $g_{\rm mix}$ appearing in the the covariant derivative,
\[D_{\mu} \varphi \,=\,
\partial_{\mu} \varphi \,+\, i  \sqrt{\frac{3}{5}}g_1 Q^Y A^Y_{\mu}  \,+\, i (g_{\rm mix} Q^Y + e_{\sst \mathrm{CW}} Q^{\sst \mathrm{CW}})A^{\sst \mathrm{CW}}_{\mu}\,.
\]
Kinetic mixing is induced radiatively in so far as there are matter fields transforming under both
Abelian factors. In the present model it is induced by the mass eigenstates of the scalar fields. In what follows for simplicity we will choose  $g_{\rm mix}(\mu=M_t)=0$ at the top mass.

%%%%%%%%%%%%%%%%%%%%%%%%%%%%%%%%%%%%%%%%%%%%%%%%%%%%
\subsection{Standard Model \texorpdfstring{$\times$~U(1)${_{\bf \sst  B-L}}$}{xU(1)B-L}}
%%%%%%%%%%%%%%%%%%%%%%%%%%%%%%%%%%%%%%%%%%%%%%%%%%%%

The RG equations in the $\mathrm{B-L}$ theory are the appropriate generalisation of the equations above. These equations were
first derived in \cite{Basso:2010jm} and were also discussed recently in \cite{Iso:2012jn}.
In our conventions the RG evolution in the CSI U(1)${_{\bf \sst  B-L}}\times$ SM
theory with the
classical scalar potential \eqref{Vhphi} is determined by the set of RG equations below:
\begin{eqnarray}
(4\pi)^2 \frac{d \lambda_H}{d \log \mu}&=&
r.h.s.\,\eqref{lHU1} \label{lHBL}\\
(4\pi)^2 \frac{d \lambda_\phi}{d \log \mu} &=& 20\lambda_\phi^2 +2\lambda_{\rm P}^2
-48\lambda_\phi \, e_{\bf \sst  B-L}^2 +96 e_{\bf \sst  B-L}^4
-Tr[(y^M)^4]+8\lambda_\phi Tr[(y^M)^2]
\label{lphiBL} \\ 
(4\pi)^2 \frac{d \lambda_{\rm P}}{d \log \mu}&=&\lambda_{\rm P}\left(6y_t^2
+12\lambda_H+8\lambda_\phi -4\lambda_{\rm P}
-24e_{\bf \sst  B-L}^2
 -\frac{9}{10}g_1^2  -\frac{9}{2}g_2^2
-\frac{3}{2}g_{\rm mix}^2 \right. \nonumber\\
&&\qquad +4 Tr[(y^M)^2] \bigg)-12g_{\rm mix}^2e_{\bf \sst  B-L}^2 
\label{lPBL} \,.
\end{eqnarray}
The Yukawas for the top quark and  for 3 Majorana neutrinos are determined via
\begin{eqnarray}
(4\pi)^2 \frac{d y_t}{d \log \mu}&=&y_t\left(\frac{9}{2}y_t^2 -\frac{17}{20}g_1^2-\frac{9}{4}g_2^2-8g_3^2
-\frac{17}{12}g_{\rm mix}^2 -\frac{2}{3}e_{\bf \sst  B-L}^2-\frac{5}{3}g_{\rm mix}e_{\bf \sst  B-L}
\right)  \label{ytBL} \\
(4\pi)^2 \frac{d y_i^M}{d \log \mu}&=&y_i^M\left(4(y_i^M)^2+Tr[(y^M)^2]-6e_{\bf \sst  B-L}^2\right)
\label{ymBL}\,,
\end{eqnarray}
and the gauge couplings are given by eqs.~\eqref{gaugeU1} together with
\begin{eqnarray}
(4\pi)^2 \frac{d e_{\bf \sst  B-L}}{d \log \mu} &=&12e_{\bf \sst  B-L}^3+\frac{32}{3}e_{\bf \sst  B-L}^2\,g_{\rm mix}
+\frac{41}{6}e_{\bf \sst  B-L}\,g_{\rm mix}^2
\label{eBL}\\
(4\pi)^2 \frac{d g_{\rm mix}}{d \log \mu} &=&
\frac{41}{6}g_{\rm mix}\left(g_{\rm mix}^2+\frac{6}{5} g_1^2\right)
+2\frac{16}{3}e_{\bf \sst  B-L}\left(g_{\rm mix}^2+\frac{3}{5}g_1^2\right)
+12e_{\bf \sst  B-L}^2\,g_{\rm mix}
\label{gmixBL}\,.
\end{eqnarray}

%%%%%%%%%%%%%%%%%%%%%%%%%%%%%%%%%%%%%%%%%%%%%%%%%%%%
\subsection{Standard Model \texorpdfstring{$\times$ U(1)${_{\bf \sst  B-L}}$ $\oplus$}{xU(1) B-L plus} real scalar}
%%%%%%%%%%%%%%%%%%%%%%%%%%%%%%%%%%%%%%%%%%%%%%%%%%%%

When discussing the Higgs vacuum stability we will soon find out that the size of the available region on the
CSI ESM parameter space will be significantly dependent on whether or not the theory includes an
additional singlet field. We are thus led to extend the RG equations above to the case with the singlet.

The scalar self-couplings and portal couplings in this model are governed by the following equations,
\begin{eqnarray}
(4\pi)^2 \frac{d \lambda_H}{d \log \mu}&=&
r.h.s.\,\eqref{lHBL}    \,+\, \frac{1}{2}\lambda^2_{H s}
 \label{lHBLs}\\
(4\pi)^2 \frac{d \lambda_\phi}{d \log \mu} &=& 
r.h.s.\,\eqref{lphiBL}    \,+\, \frac{1}{2}\lambda^2_{\phi s}
\label{lphiBLs} \\ 
(4\pi)^2 \frac{d \lambda_{\rm P}}{d \log \mu}&=&
r.h.s.\,\eqref{lPBL}   \,-\, \lambda_{H s}\lambda_{\phi s}
\label{lPBLs} \\
(4\pi)^2 \frac{d \lambda_{s}}{d \log \mu} &=& 18\lambda_s^2+\lambda_{\phi s}^2+2\lambda_{H s}^2
\label{lsBLs}\\
(4\pi)^2 \frac{d \lambda_{Hs}}{d \log \mu} &=&
\lambda_{Hs}\left(6y_t^2+12\lambda_H+ 6\lambda_s + 4 \lambda_{Hs}
-\frac{9g_1^2}{10}-\frac{9g_2^2}{2}
\right) -2\lambda_{\rm P}\lambda_{\phi s}
\label{lHsBLs} \\
(4\pi)^2 \frac{d \lambda_{\phi s}}{d \log \mu}&=&
\lambda_{\phi s}\left(12\lambda_{\phi}+ 6\lambda_s
+4 \lambda_{\phi s} -18e_{\bf \sst  B-L}^2 \right)  -4\lambda_{\rm P}\lambda_{H s}
\,.
\label{lphisBLs}
\end{eqnarray}

The rest of the RG equations are the same as before.
Equations for Yukawa couplings are \eqref{ytBL}-\eqref{ymBL}, 
and the gauge couplings are given by eqs.~\eqref{gaugeU1} together with \eqref{eBL}-\eqref{gmixBL}.
As always, we set $g_{\rm mix}(\mu=M_t)=0$.

Note that it is easy to derive a simple formula,  eq.~\eqref{eq:simple} below, which computes the
coefficients in front of scalar couplings on the right hand sides of the RG equations.
First, let us write the classical scalar potential in the form,
\[
V_0 = \sum_{\varphi} \,\frac{\lambda_{\varphi}}{4}( \vec\varphi^{\,2})^2 \,+\,
\sum_{\varphi < \varphi'} \,\frac{\lambda_{\varphi \varphi'}}{4}( \vec\varphi^{\,2})( \vec\varphi^{\,\prime\,2})\,,
\]
where in our case $\varphi =\{h, \phi, s\}$, and the second sum is understood as over the three pairs of indices,
$(h,\phi)$, $(h,s)$ and $(\phi,s)$. The notation $\vec \varphi$ denotes the canonically normalised real components of the 
Higgs, $\vec h= (h_1,\ldots, h_4)$, the complex doublet $\vec \phi= (\phi_1,\ldots, \phi_4)$ and the
real singlet $\vec s= s$. In general we denote the number of real components of each of the species of
$\vec \varphi$ and $N_{\varphi}$.
It is then easy to derive the expressions for scalar-coupling contributions to all the self-interactions, by
counting the contributing 4-point 1PI diagrams involving 2 scalar vertices. For the beta functions of the self-couplings we get,
\[
(4\pi)^2 \frac{d \lambda_{\varphi}}{d \log \mu}\,\ni\,
2 (N_{\varphi}+8)\, \lambda_{\varphi}^2 \,+\,
\sum_{\tilde{\varphi}} \frac{N_{\tilde{\varphi}}}{2} \,\lambda^2_{\varphi \tilde{\varphi}}\,,
\]
and the portal couplings are governed by,
\[
(4\pi)^2 \frac{d \lambda_{\varphi \varphi'}}{d \log \mu}\,\ni\,
\sum_{\varphi} 2 (N_{\varphi}+2)\,  \lambda_{\varphi}\lambda_{\varphi \varphi'}
\,+\,
\sum_{\varphi'} 2 (N_{\varphi'}+2)\,\lambda_{\varphi \varphi'} \lambda_{\varphi'}
\,+\,
\sum_{\tilde{\varphi}} N_{\tilde{\varphi}}\,\lambda_{\varphi \tilde\varphi}\lambda_{\varphi' \tilde\varphi}
\,+\,
4\,\lambda^2_{\varphi \varphi'}
\label{eq:simple}
\]
This formula is valid for all of the CSI ESM examples considered in this paper.

%%%%%%%%%%%%%%%%%%%%%%%%%%%%%%%%%%%%%%%%%%%%%%%%%%%%
\subsection{Standard Model \texorpdfstring{$\times$ SU(2)$_{\bf \sst \mathrm{CW}}$}{x SU(2) CW}}
%%%%%%%%%%%%%%%%%%%%%%%%%%%%%%%%%%%%%%%%%%%%%%%%%%%%

We can also write down the relevant renormalisation group equations for the classically scale-invariant Standard Model $\times$ SU(2)$_{\sst \mathrm{CW}}$ theory with the scalar potential given by eq.~\eqref{VSU2}. These RG equations were first derived  in Refs.~\cite{Hambye:2013dgv,Carone:2013wla}. For scalar self-couplings $\lambda_H$ and $\lambda_\phi$, and the portal coupling $\lambda_{\rm P}$ we have:
\begin{eqnarray}
&&(4\pi)^2 \frac{d \lambda_H}{d \log \mu}=-6 y_t^4+24\lambda_H^2+2\lambda_{\rm P}^2 +
\lambda_H \left(12y_t^2-\frac{9}{5}g_1^2-9g_2^2\right) +\frac{27}{200}g_1^4+\frac{9}{20}g_2^2g_1^2
+\frac{9}{8}g_2^4\qquad \quad
\label{lHeqn}
% \nonumber\\ 
\\
&&(4\pi)^2 \frac{d \lambda_\phi}{d \log \mu}= 24\lambda_\phi^2 +2\lambda_{\rm P}^2
-9\lambda_\phi \, g_{\sst \mathrm{CW}}^2 +\frac{9}{8} g_{\sst \mathrm{CW}}^4\\ 
&&(4\pi)^2 \frac{d \lambda_{\rm P}}{d \log \mu}=\lambda_{\rm P}\left(6y_t^2
+12\lambda_H+12\lambda_\phi
-4\lambda_{\rm P}
-\frac{9}{2}g_{\sst \mathrm{CW}}^2-\frac{9}{10}g_1^2-\frac{9}{2}g_2^2\right)\,,
\end{eqnarray}
where the top Yukawa coupling obeys
\begin{eqnarray}
(4\pi)^2 \frac{d y_t}{d \log \mu}=y_t\left(\frac{9}{2}y_t^2 -\frac{17}{20}g_1^2-\frac{9}{4}g_2^2-8g_3^2\right)\,,
\end{eqnarray}
and $g_{\sst \mathrm{CW}}$, $g_{3,2,1}$ are the gauge couplings of the SU(2)$_{\sst \mathrm{CW}} \times$ SU(3) $\times$ SU(2) $\times$ U(1),
\begin{eqnarray}
&&(4\pi)^2 \frac{d g_{\sst \mathrm{CW}}}{d \log \mu}=-\frac{43}{6}g_{\sst \mathrm{CW}}^3-\frac{1}{(4\pi)^2}\frac{259}{6}g_{\sst \mathrm{CW}}^5\\
\nonumber\\
&&(4\pi)^2 \frac{d g_3}{d \log \mu}=-7g_3^3 \,,\quad
(4\pi)^2 \frac{d g_2}{d \log \mu}=-\frac{19}{6}g_2^3  \,,\quad
(4\pi)^2 \frac{d g_1}{d \log \mu}=\frac{41}{10}g_1^3\,,
\end{eqnarray}
where for the U(1) coupling we use the 
normalisation $g_1^2 = \frac{5}{3} g_Y^2$.

All running couplings are computed in the $\overline{\rm MS}$ scheme and furthermore we use the physical freeze-out
condition for the SU(2)$_{\sst \mathrm{CW}}$ degrees of freedom at the RG scales below their mass shell. In other words,
the SU(2)$_{\sst \mathrm{CW}}$ contributions to the $\beta$-functions for $g_{\sst \mathrm{CW}}$, $\lambda_\phi$ and $\lambda_{\rm P}$
will be set to zero when 
$\mu<M_{Z'}=\frac{1}{2}g_{\sst \mathrm{CW}} \langle\phi\rangle.$

%%%%%%%%%%%%%%%%%%%%%%%%%%%%%%%%%%%%%%%%%%%%%%%%%%%%
\subsection{Standard Model \texorpdfstring{$\times$ SU(2)$_{\bf \sst \mathrm{CW}}$ $\oplus$}{x SU(2)CWplus} real scalar}
%%%%%%%%%%%%%%%%%%%%%%%%%%%%%%%%%%%%%%%%%%%%%%%%%%%%

RG-equations for the three scalar self-couplings now take the form:
\begin{eqnarray}
(4\pi)^2 \frac{d \lambda_H}{d \log \mu}&=&-6 y_t^4+24\lambda_H^2+2\lambda_{\rm P}^2 
+ \frac{1}{2}\lambda_{Hs}^2 
\label{lHSU2s}
\nonumber\\
&&+ \lambda_H \left(12y_t^2-\frac{9}{5}g_1^2-9g_2^2\right) +\frac{27}{200}g_1^4+\frac{9}{20}g_2^2g_1^2
+\frac{9}{8}g_2^4\\
(4\pi)^2 \frac{d \lambda_\phi}{d \log \mu} &=& 24\lambda_\phi^2 +2\lambda_{\rm P}^2 + \frac{1}{2}\lambda_{\phi s}^2
-9\lambda_\phi \, g_{\sst \mathrm{CW}}^2 +\frac{9}{8} g_{\sst \mathrm{CW}}^4\\ 
(4\pi)^2 \frac{d \lambda_{s}}{d \log \mu}&=&
18 \lambda_s^2+2\lambda_{\phi s}^2+2\lambda_{H s}^2\,,
\end{eqnarray}
and for the three portal couplings we have,
\begin{eqnarray}
(4\pi)^2 \frac{d \lambda_{\rm P}}{d \log \mu}&=&\lambda_{\rm P}\left(6y_t^2
+12\lambda_H+12\lambda_\phi
-4\lambda_{\rm P}
-\frac{9}{2}g_{\sst \mathrm{CW}}^2-\frac{9}{10}g_1^2-\frac{9}{2}g_2^2\right)
- \lambda_{Hs}\lambda_{\phi s} \qquad \quad\\
(4\pi)^2 \frac{d \lambda_{Hs}}{d \log \mu}&=&\lambda_{Hs}\left(6y_t^2 +12\lambda_H+ 6\lambda_s
+4\lambda_{\rm Hs}
-\frac{9}{10}g_1^2
-\frac{9}{2}g_2^2\right)  - 4\lambda_{\rm P}\lambda_{\phi s}\\
(4\pi)^2 \frac{d \lambda_{\phi s}}{d \log \mu}&=&\lambda_{\phi s}\left(12\lambda_{\phi}+ 6\lambda_s
+4\lambda_{\rm \phi s}
-\frac{9}{2}g_{\sst \mathrm{CW}}^2 \right) - 4\lambda_{\rm P}\lambda_{H s}\,.
\end{eqnarray}

%%%%%%%%%%%%%%%%%%%%%%%%%%%%%%%%%%%%%%%%%%%%%%%%%%%%
\subsection{Initial conditions and stability bounds}
%%%%%%%%%%%%%%%%%%%%%%%%%%%%%%%%%%%%%%%%%%%%%%%%%%%%

To solve the RG equations and determine the RG evolution of the couplings of our models, we first 
need to specify the initial conditions for all the couplings. 

First, 
we specify the initial conditions for the SM coupling constants at $M_t$:
the top Yukawa coupling $y_t$ 
and the SM gauge couplings initial values are taken from Ref.~\cite{Buttazzo:2013uya},
\begin{eqnarray}
y_t(\mu=M_t) &=& 0.93558 +0.00550 \left( \frac{M_t}{\rm GeV}-173.1\right)+ 
\nonumber\\
 &&-0.00042 \frac{\alpha_3(M_z)-0.1184}{0.0007}  -0.00042\frac{M_W-80.384 {\rm GeV}}{\rm GeV} \pm{0.00050}_{\rm th}
 \\
 \nonumber\\
g_3(\mu=M_t)  &=& 1.1666 +0.00314\frac{\alpha_3(M_z)-0.1184}{0.0007} -0.00046 \left( \frac{M_t}{\rm GeV}-173.1 \right)
\\
 \nonumber\\
g_2(\mu=M_t) &=& 0.64822 +0.00004 \left( \frac{M_t}{\rm GeV}-173.1 \right)+ 0.00011 \frac{M_W-80.384 {\rm GeV}}{\rm GeV}
\\
 \nonumber\\
 g_1(\mu=M_t) &=& \sqrt{\frac{5}{3}}\left(0.35761 +0.00011 \left( \frac{M_t}{\rm GeV}-173.1 \right)- 0.00021
\frac{M_W-80.384{\rm GeV}}{\rm GeV}\right)\,.
 \nonumber\\
\end{eqnarray}
In our numerical analysis we will always assume the central values for $M_t$ and $M_W$.

The CW portal coupling, $\lambda_{\rm P}$ and the CW gauge coupling are taken as the two free input parameters
specifying the 2-dimensional BSM parameter space of our U(1) or SU(2) $\times$ SM theories. When an additional
singlet field $s(x)$ is present, the input parameters also include  $\lambda_{Hs}$, $\lambda_{s}$ and $\lambda_{\phi s}$.

The input values of the two remaining couplings, the Higgs self-coupling $\lambda_H$, and the self-coupling of the
CW scalar, $\lambda_\phi$, are then determined from the value of the SM Higgs mass, and from the CW matching condition
\eqref{eq:cwmsbar-P}, respectively. 
To find $\lambda_H$ we numerically compute the eigenvalues of the mass matrix \eqref{Mmixing} 
and set $m_{h_1}=125.66$ GeV, as was outlined in eq.~\eqref{SMh1-ident}. 
We then iteratively solve for $\lambda_\phi(\mu=M_t)$ by running it from the top mass scale to $\mu=\left<\phi\right>$ and checking that we fulfil the CW matching relation \eqref{eq:cwmsbar-P} at the latter scale. 

Having thus specified the initial conditions for all couplings at the low scale, $\mu=M_t$, we
 run them up to the high scale $\mu=M_{\rm Pl}$ by numerically solving the RG equations. To determine the region of the parameter space where the Higgs potential is stable,
we check that the conditions,
\[
4\lambda_H (\mu)\, \lambda_\phi  (\mu)\,>\, \lambda_{\rm P}^2 (\mu)\,, \qquad \lambda_H (\mu)\,>\,0\,, 
\quad {\rm for \,\, all} \,\, \mu\le M_{\rm Pl}\,,
\label{treelst}
\]
arising from the positive definiteness of eq.~\eqref{Vhphi}
are fulfilled. 
We also check that the model remains perturbative, requiring that all its scalar couplings are bounded by an order-1 constant 
all the way to the Plank scale,
\[\lambda_i (\mu)\, <\, {\rm const}\,  {\cal O}(1) \,=3\,,\]
where for concreteness we chose a conservative numerical value of the upper bound = 3; in practice our results do not depend
significantly on this choice.

\begin{figure}[t!]
\centering
\includegraphics[width=0.7\columnwidth]{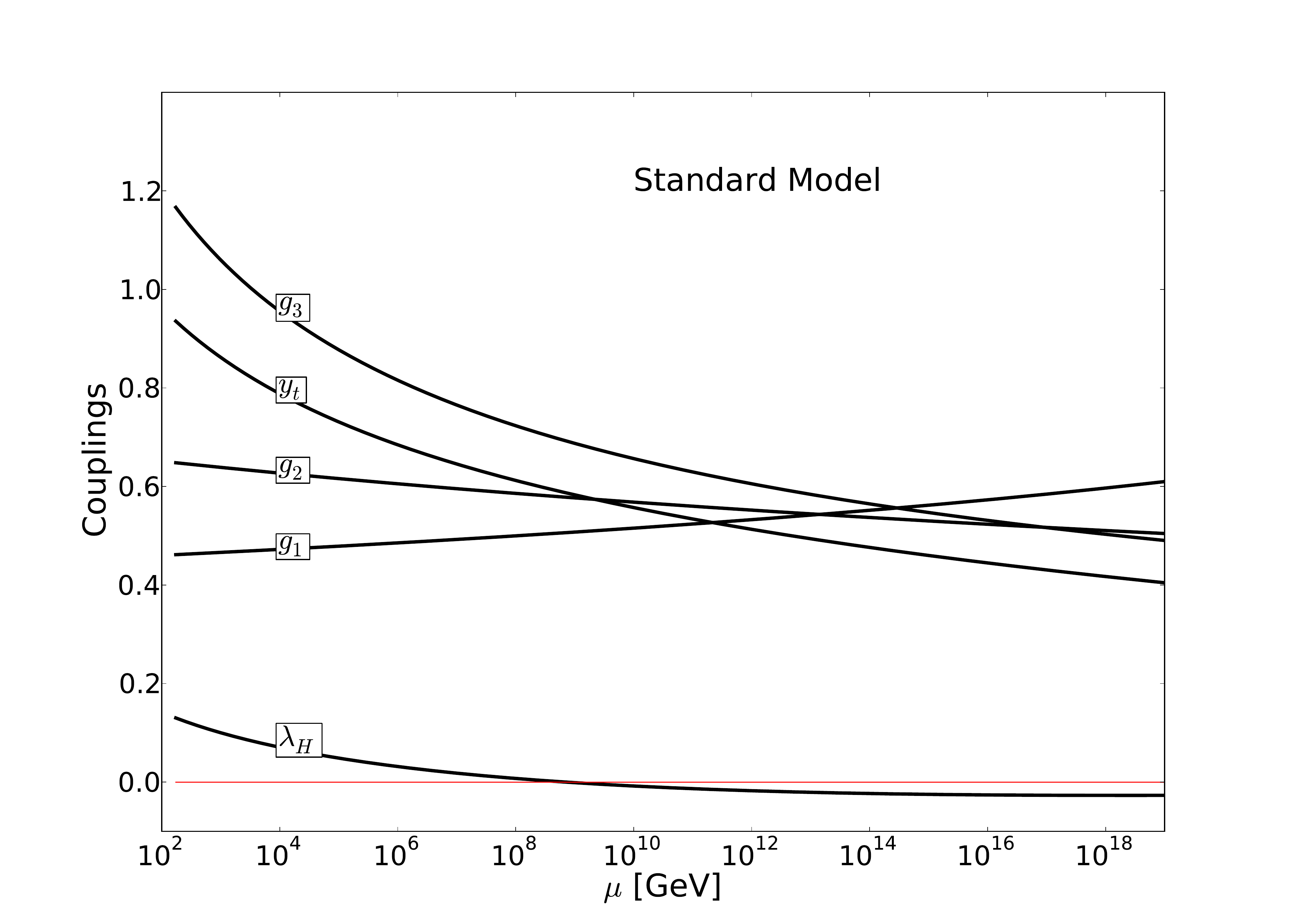}
\caption{RG evolution in the Standard Model. The Higgs self-coupling turns negative at \mbox{$\mu \gtrsim 10^9$~GeV} thus signalling
that the SM Higgs potential becomes unstable below the Planck scale. In this and all other Figures we use $M_t =173.1$ GeV.}
\label{Ffig:rge_sm}
\end{figure}

%%%%%%%%%%%%%%%%%%%%%%%%%%%%%%%%%%%%%%%%%%%%%%%%%%%%
%%%%%%%%%%%%%%%%%%%%%%%%%%%%%%%%%%%%%%%%%%%%%%%%%%%%%%%%%%%%
\section{Higgs Physics: stability and phenomenology}
\label{sec:Higgs}
%%%%%%%%%%%%%%%%%%%%%%%%%%%%%%%%%%%%%%%%%%%%%%%%%%%%%%%%%%%%
%%%%%%%%%%%%%%%%%%%%%%%%%%%%%%%%%%%%%%%%%%%%%%%%%%%%

\begin{figure}[t!]
\centering
\includegraphics[width=0.61\columnwidth]{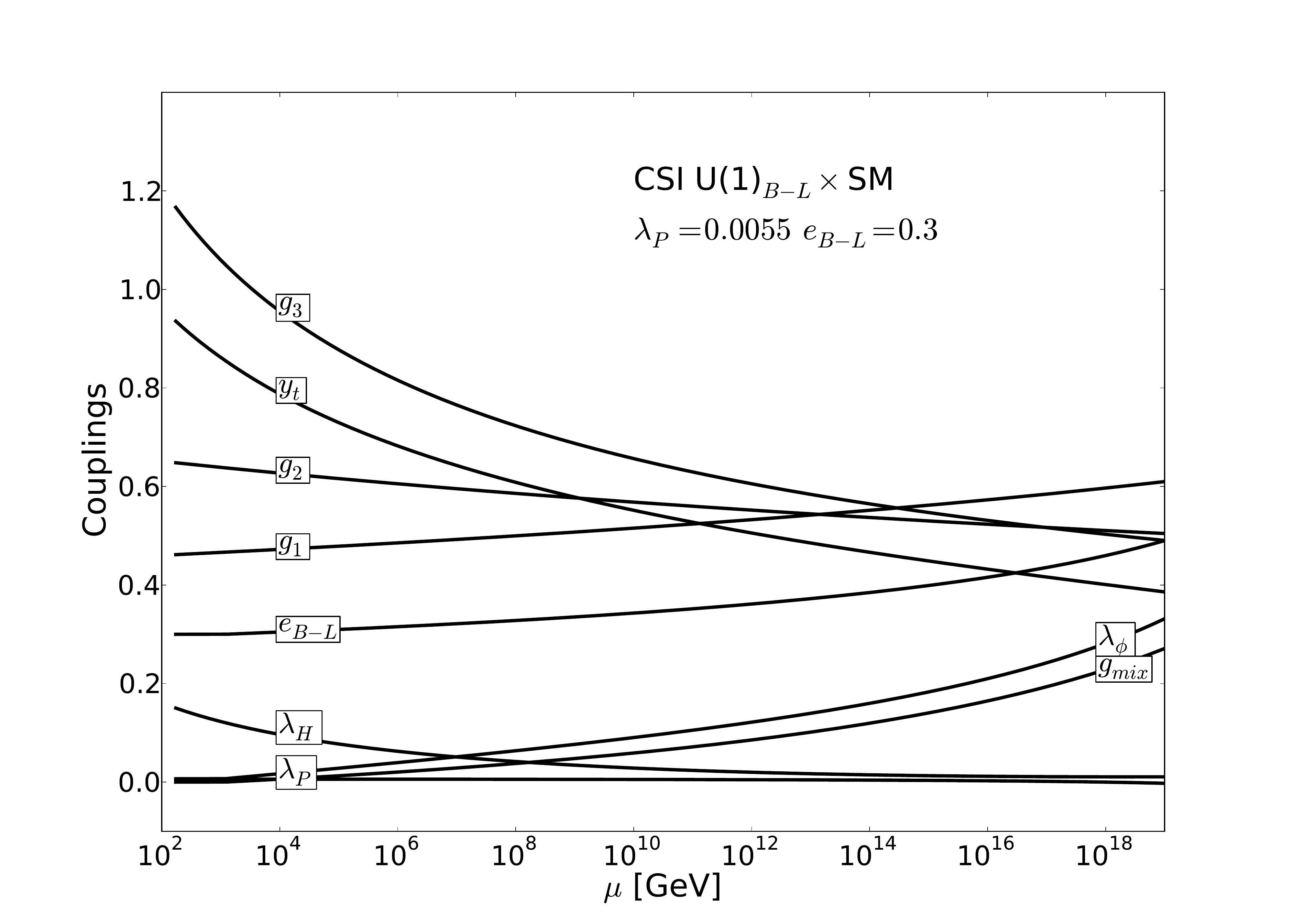}\\
\includegraphics[width=0.61\columnwidth]{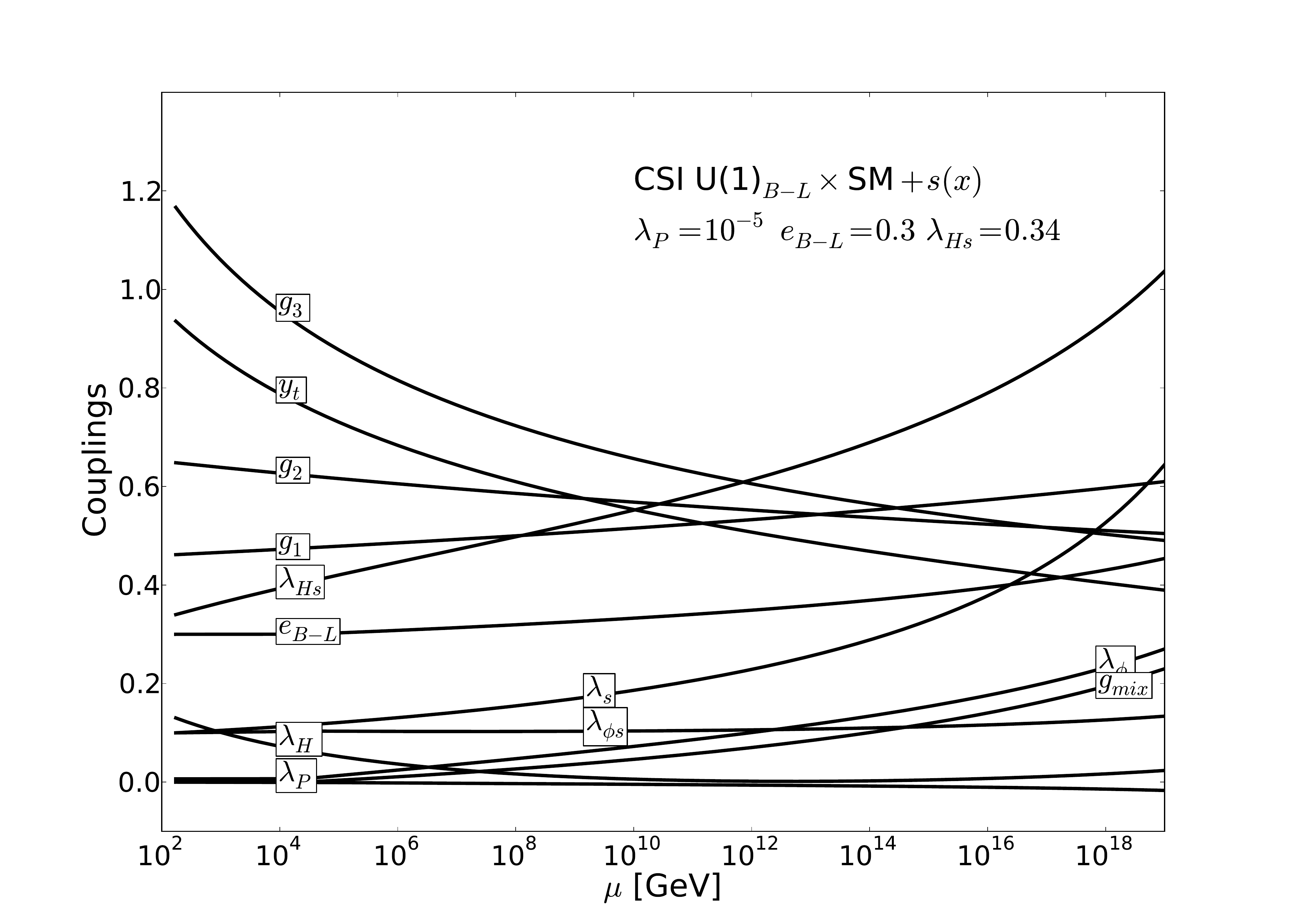}\\
\includegraphics[width=0.61\columnwidth]{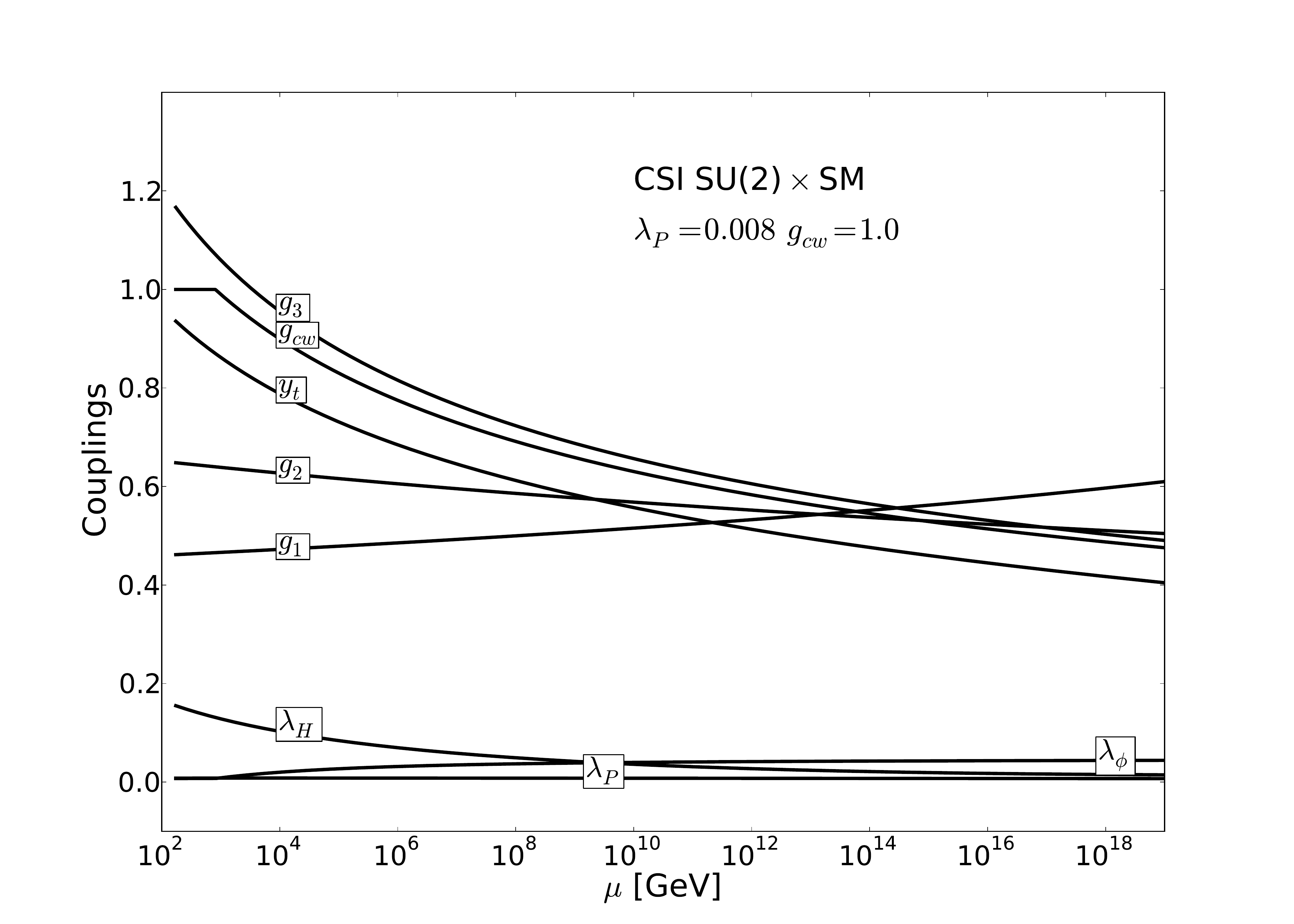}
\caption{\noindent RG evolution in CSI $\mathrm{E}$SM theories with  (a) $\mathrm{E}= \mathrm{U}(1)_{\bf \sst  B-L}$, (b)  $\mathrm{E}=\mathrm{U}(1)_{\bf \sst  B-L} + s(x),$
and (c) $\mathrm{E}=\mathrm{SU}(2)_{\sst \mathrm{CW}}$. With these initial conditions the Higgs coupling $\lambda_H$ stays positive 
and satisfies the tree-level stability bound \eqref{treelst}.}\label{fig:3runs}
\end{figure}

%%%%%%%%%%%%%%%%%%%%%%%%%%%%%%%%%%%%%%%%
\begin{figure}[t!]
\centering
\includegraphics[width=.75\columnwidth]{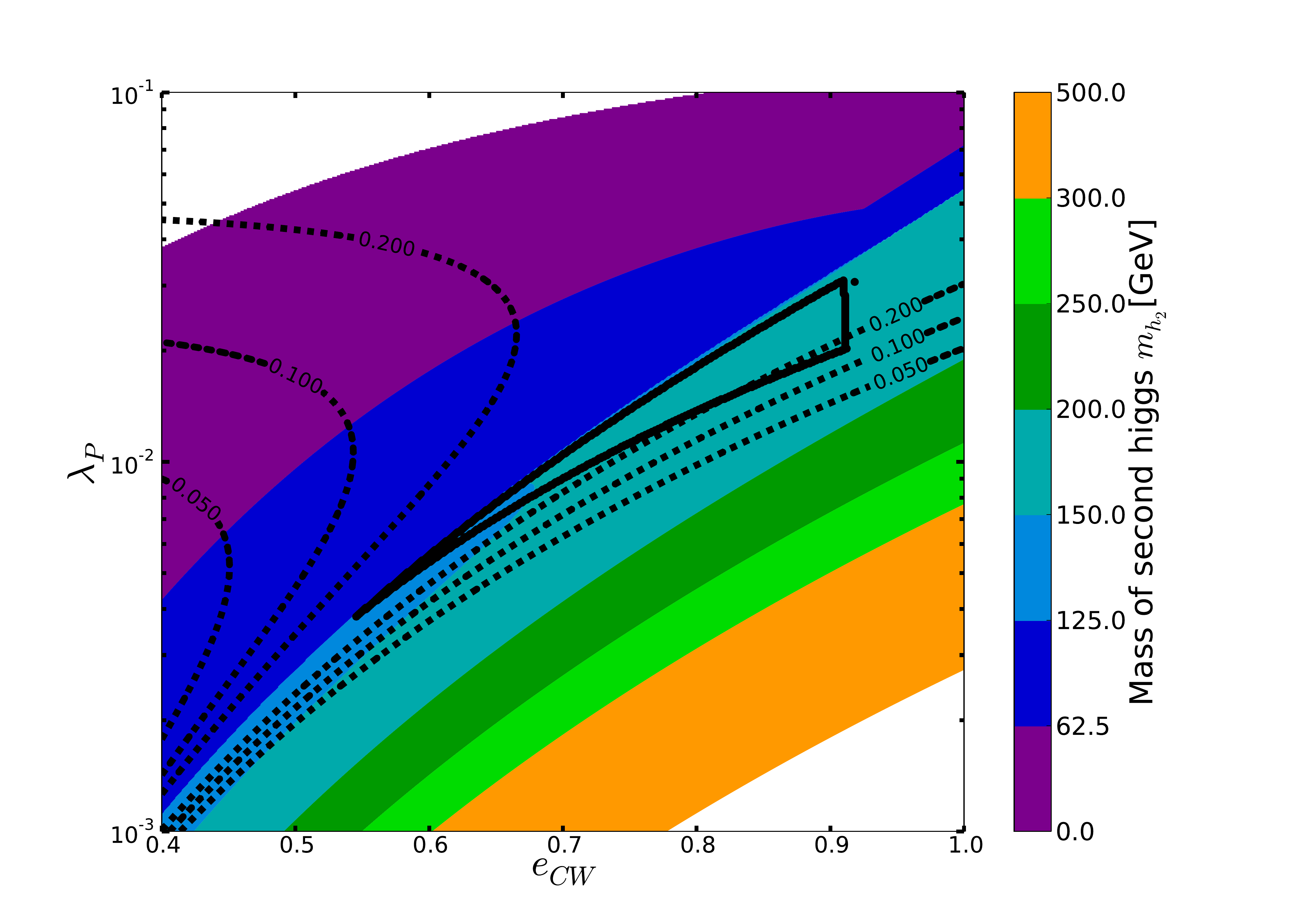}
\caption{Parameter space in the minimal
 U(1)$_{\sst \mathrm{CW}}\,\times$ \!SM classically scale invariant theory.
The black wedge-shaped contour shows the region of the $(\lambda_{\rm P}, e_{\sst \mathrm{CW}})$ parameter space of the model
where the Higgs potential is stabilised. The dotted lines represent contours of fixed values
 $\sin^2 \theta = $ 0.05, 0.1 and 0.2 
  of the Higgs mixing angle. Finally, the colour-coding indicates the mass of the second scalar $h_2$ in GeV.}
\label{U1Stab}
\end{figure}
%%%%%%%%%%%%%%%%%%%%%%%%%%%%%%%%%%%%%%%%

It is well known that in the Standard Model the Higgs self-coupling becomes negative at \mbox{$\mu \sim 10^9$~GeV}
making the SM Higgs potential unstable below the Planck scale \cite{Degrassi:2012ry,Buttazzo:2013uya}
(see also~\cite{Sher:1988mj,Lindner:1985uk} for a review of earlier work). This effect can be seen in fig.~\ref{Ffig:rge_sm}
which shows the solution of RG equations in the limit where all Higgs portal interactions are switched off.

For our classically scale invariant extensions of the SM to be meaningful and practical natural theories 
valid all the way up to the Planck scale, the Higgs potential has to be 
stabilised.\footnote{In this paper we will concentrate on the more conservative case of absolute 
stability.
Another phenomenologically acceptable possibility analysed recently in \cite{Buttazzo:2013uya}
is that the SM vacuum is metastable, with a lifetime much greater 
than the age of the Universe.
In that case one would also have to argue why after reheating the Universe ended up in the metastable vacuum near the 
origin, for example following the approach of \cite{Abel:2006cr}.}
There are two mechanisms, both relying on the Higgs portal interactions, to achieve this:
\begin{enumerate}
\item The SM Higgs is the mixed mass eigenstate $h_1$ between $H$ and the CW scalar as dictated by
eq.~\eqref{SMh1-ident}. As we explained at the end of section {\bf \ref{sec2:U1CW}} in the case
where the second scalar is heavier than the Higgs, $m_{h_2} > m_{h_1}$,  
the initial value of the Higgs self-coupling $\lambda_H$ is larger than in the SM, cf.~eq.~\eqref{lSM-},
 and this helps with the Higgs stabilisation \cite{Lebedev:2012zw,EliasMiro:2012ay,Hambye:2013dgv}.
 
\item The portal couplings of other scalars to the Higgs, such as $\lambda_{P}$ and $\lambda_{Hs}$
contribute positively to the beta function of $\lambda_H$ as can be seen e.g. from the RG equation \eqref{lHSU2s}
in the SU(2)$_{\sst \mathrm{CW}}$ + scalar case, where $\beta_{\lambda_H} \ni 2\lambda_{\rm P}^2 
+ \frac{1}{2}\lambda_{Hs}^2$. This effect (in particular due to the otherwise unconstrained but still perturbative
$\lambda_{Hs}$ coupling) will be instrumental in achieving the Higgs stability in models with an extra scalar,
\cite{Gonderinger:2009jp,Lebedev:2011aq}.
\end{enumerate}
Examples of RG running for some specific input values of parameters for three different classes of models 
which result in stable Higgs potential are shown in fig.~\ref{fig:3runs} where cases (a) and (c) give an example of mechanism
(1.) and the model with an additional scalar in case (b) is a representative of mechanism (2.) at work.

In the rest of this section we will quantify the regions of the parameter spaces for individual
models where the scalar potential is stabilised.
 We will also combine these considerations with the current LHC limits applied to the extended Higgs sectors of our 
Higgs portal theories in a model by model basis.

%%%%%%%%%%%%%%%%%%%%%%%%%%%%%%%%%%%%%%%%%%%%%%%%%%%%
\subsection{CSI \texorpdfstring{U(1)${_{\bf \sst  CW}}\,\times$ \!SM}{U(1)CWxSM}}
\label{sec:4.1}
%%%%%%%%%%%%%%%%%%%%%%%%%%%%%%%%%%%%%%%%%%%%%%%%%%%%

In this theory the mechanism (1.) is operational for stabilising the Higgs potential in a region of the 2-dimensional
parameter space of the model described by $\lambda_{\rm P}$ and the CW gauge coupling.
As shown in fig.~\ref{U1Stab} we get a wedge shaped region inside the black contour inside which the Higgs potential is stable.

Higgs stabilisation in this region can be traced to the initial value of $\lambda_H$ being enhanced compared to the SM
 due to mixing between $h$ and the CW scalar field. The wedge shape can be understood as follows. The upper edge of the wedge follows the mass contour where $m_{h_2}>m_h$ since the enhancement of the initial value of $\lambda_h$ only happens when $m_{h_2}>m_{h_1}$, see~\eqref{lSM-}. The mechanism is only effective when the two masses are not too far from each other (cf.~the denominator of the second term in eq.~\eqref{lSM-}). The lower contour of the wedge signifies when the mass difference becomes too large.  The effect is enhanced when the off-diagonal element is larger as we get more mixing. This explains why the stability wedge in fig.~\ref{U1Stab} is wider for larger values of~$\lambda_{\rm P}$. We get an upper limit on~$e_{\sst CW}\approx 0.9 $ since for larger values we find a Landau pole before the Planck scale. 

Higgs sector phenomenology of this model 
 in the context of LHC and LEP, future
colliders and low energy measurements was analysed recently in \cite{Englert:2013gz}.
In particular, it was shown there that on the part of the parameter space where the second scalar is light, 
 \mbox{$10^{-4}$ GeV $<\, m_{h_2}\,<\, m_{h_1}/2$}
the presently available Higgs data (and specifically the limits on the invisible Higgs decays)
constrain the model quite tightly 
by placing the upper limit on the 
portal coupling to be $\lambda_{\rm P} \lesssim 10^{-5}$.

However, from fig.~\ref{U1Stab} we see that the Higgs stability in the minimal model (and more generally in all
portal models without additional scalar $s(x)$, i.e. relying on the stabilisation mechanism (1.) )
requires the second scalar to be heavier than the SM Higgs, $m_{h_2} \,>\, m_{h_1}$ 
(see also figs.~\ref{U1BLStab}, \ref{SU2Stab}). Thus Higgs stability 
pushes these models in to the region of the parameter space with the heavier second scalar, precisely where
the collider limits on invisible Higgs decays and on non-observation of other Higgs-like states are much less stringent.

Collider limits which do constrain the stability region in fig.~\ref{U1Stab} are the exclusion limits on the
heavier Higgs production normalised to the expected SM cross-section at this Higgs mass. In all Higgs portal models we consider in this paper, the expected cross-section for the $h_2$ scalar is given by the SM cross-section times $\sin^2 \theta$ of the mixing
angle. With the currently available ATLAS and CMS data for the search of the heavier Higgs boson
at integrated luminosity of up to 5.1 fb$^{-1}$ at $\sqrt{s}=7$ TeV and up to 5.3  fb$^{-1}$ at $\sqrt{s}=8$ TeV,
the observed signal 
strength in the units of the SM cross-section for the heavier Higgs is roughly at the level of $10^{-1}$, or slightly above,
as can be seen from plots in \cite{ATLASsc, CMSWWZZ, CMSZZ4l}.
This gives an upper limit on the mixing angle 
$\sin^2 \theta \lesssim 0.1$. 

The contours of constant values of $\sin^2 \theta =0.05,\, 0.1$ and 0.2 are shown on fig.~\ref{U1Stab} as dotted lines.
As we can see for $\sin^2 \theta  \lesssim 0.1$  there is no overlap left between 
what is allowed by the collider limits and what is consistent with the Higgs stability in this model. We thus conclude that 
the combination of the Higgs potential stabilisation and the LHC limits on the heavier Higgs essentially rule out the 
minimal U(1)$_{\sst \mathrm{CW}}\times$ SM theory. This conclusion is based on the one-loop RG analysis, on the methodology we adopted for the selection of initial values,
and on the use of the central value for the top mass. As such there is an intrinsic theoretical uncertainty in the exact position of the wedge.
By lowering the top mass from its central value by 1 GeV, the wedge in fig.~\ref{U1Stab} would touch the
$\sin^2 \theta  = 0.1$ contour making the model viable in the limited corner of the parameter space.

Instead, to get a stable viable model with the current central value of the top mass and without relying upon the sub-leading
RG effects, we will simply extend the theory by adding a singlet $s(x)$ in sections {\bf \ref{sec:4.3}}, {\bf \ref{sec:4.5}}.

%%%%%%%%%%%%%%%%%%%%%%%%%%%%%%%%%%%%%%%%%%%%%%%%%%%%
\subsection{CSI \texorpdfstring{U(1)${_{\bf \sst  B-L}}\,\times$ \!SM}{U(1) B-LxSM}}
\label{sec:4.2}
%%%%%%%%%%%%%%%%%%%%%%%%%%%%%%%%%%%%%%%%

\begin{figure}[]
\centering
\includegraphics[width=.75\columnwidth]{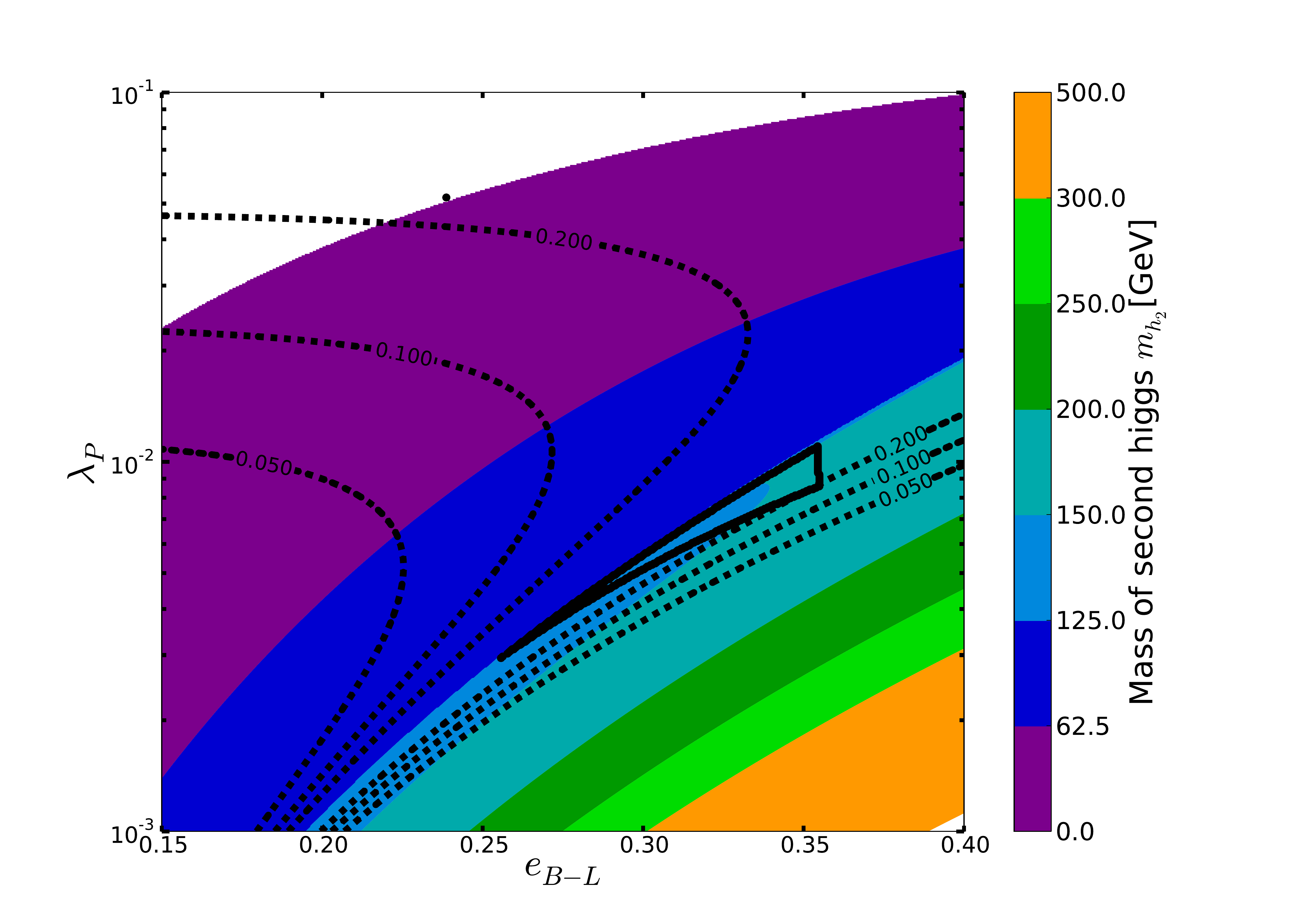}
\caption{Parameter space of the
 U(1)$_{\bf \sst  B-L}\times$ SM theory showing the region
where the Higgs potential is stabilised and the 
 $\sin^2 \theta$ contours.
The legend is the same as in Figure \ref{U1Stab}.}
\label{U1BLStab}
\end{figure}

One way to extend the minimal model is 
to allow for interactions of the hidden sector with the SM fermions. As we have seen already, a simple implementation of this 
idea is described  by the U(1)$_{\bf \sst  B-L}\times \mathrm{SM}$ classically scale invariant theory. We proceed to solve the RG equations in this 
model  and search for the region on the parameter space 
where the scalar potential is stable, with the results shown in fig.~\ref{U1BLStab}.

The stability region in fig.~\ref{U1BLStab} is shorter along the horizontal $e_{\bf \sst  B-L}$--direction than 
in the minimal CW model of fig.~\ref{U1Stab} before. This is caused by the
slope of the $\mathrm{B-L}$ gauge coupling being steeper than for the minimal U(1)$_{\sst \mathrm{CW}}\times$ SM 
theory, due to the SM quarks and leptons
which are now charged under the U(1)$_{\bf \sst  B-L}$ gauge group. 
We therefore get a Landau pole before the Planck Scale if $e_{\bf \sst  B-L}(\mu=m_t)\gtrsim 0.35$, and this 
shortens the allowed region.

The width of the stability wedge reflects the fact that in the $\mathrm{B-L}$ model the CW scalar $\phi$ has the charge of two.
As the result one would expect that the width of the $\mathrm{B-L}$ model stability region for a fixed value of the gauge coupling,
say at  $e_{\bf \sst  B-L}=0.3$, should be of similar size to the case of the pure U(1) CW sector at the twice the value of the coupling,
i.e. at $e_{\sst \mathrm{CW}}=0.6$, which is indeed the case.

Collider exclusion limits of $\sin^2 \theta \lesssim 0.1$ are indicated in  fig.~\ref{U1BLStab}  as before by the dotted lines
showing contours of constant $\sin^2 \theta =0.05,\, 0.1$ and 0.2.
We see that 
the combination of the Higgs potential stabilisation and the LHC limits on the heavier Higgs rules out also the 
U(1)$_{\bf \sst  B-L}\times$ SM theory without an additional singlet. 

In the $U(1)_{\bf \sst  B-L}$ model we also have a $Z'$ boson which couples to the Standard Model fermions. 
The ATLAS and CMS experiments give lower limits for $M_{Z'}$ of about 3 TeV \cite{ATLAS-C017,CMS-061}. 
This implies,
\[
M_{Z'}=2e_{\bf \sst  B-L}\left<\phi\right>=2e_{\bf \sst  B-L}\sqrt{\frac{2\lambda_H}{\lambda_p}}v\,,
\]
and therefore
\[
\sqrt{\lambda_{\rm P}}\, <\,
\frac{2v\sqrt{2\lambda_H}}{3\, {\rm TeV}}e_{\bf \sst  B-L}\,\,\implies \,\,\lambda_p\lesssim (0.1\, e_{\bf \sst  B-L})^2\,.
\]
For $e_{\bf \sst  B-L}=0.35$ we find that  $\lambda_{\rm P}\lesssim 10^{-3}$, which is clearly outside
the stability wedge of the $\mathrm{B-L}$ model. Therefore Higgs stabilisation in the minimal  U(1)$_{\bf \sst  B-L}\times\mathrm{SM}$ theory
is also not compatible with the collider limits on $Z'$.

%%%%%%%%%%%%%%%%%%%%%%%%%%%%%%%%%%%%%%%%%%%%%%%%%%%%
\subsection{CSI \texorpdfstring{U(1)${_{\bf \sst  B-L}}\,\times$ \!SM $\oplus$}{U(1) B-LxSM plus} singlet}
\label{sec:4.3}
%%%%%%%%%%%%%%%%%%%%%%%%%%%%%%%%%%%%%%%%%%%%%%%%%%%%

When we add a real scalar $s(x)$ to the U(1)$_{\sst \mathrm{CW}}$ or U(1)$_{\bf \bf \sst B-L} \times$ SM theory, the scalar potential 
is stabilised by the mechanism (2.) which relies on the positive shift in the $\beta$-function for~$\lambda_H$,
\[
\beta_{\lambda_H} \,\ni\, + \frac{\lambda_{Hs}}{2}\,.
\label{bhs}
\]
We have checked that the stabilisation occurs on the entire $(\lambda_{\rm P}, e)$ 2d parameter space for values of $\lambda_{Hs} \sim 0.34$ or above, as can be seen from the 
left table in Table \ref{U1BLSStab}.
\begin{table}[t]
  \small
  \centering
 {%
    \hspace{.5cm}%
   \begin{tabular}{c c c}
 \hline\hline 
 $\lambda_{\rm P}$ & $e_{\bf \sst  B-L}$ & $\lambda_{Hs}$   \\ [0.5ex] 
 \hline % inserts single horizontal line
$10^{-5}$&0.1&0.34\\
$10^{-5}$&0.2&0.34\\
$10^{-5}$&0.3&0.33\\
0.0001&0.1&0.35\\
0.0001&0.2&0.34\\
0.0001&0.3&0.33\\
0.001&0.1&0.35\\
0.001&0.2&0.29\\
0.001&0.3&0.33\\
 \hline
 \end{tabular}
    \hspace{.5cm}%
}\hspace{1cm}
  {%
    \hspace{.5cm}%
  \begin{tabular}{c c c }
 \hline\hline 
$\lambda_{\rm P}$ & $g_{\sst \mathrm{CW}}$ & $\lambda_{Hs}$   \\ [0.5ex] 
 \hline % inserts single horizontal line
$10^{-5}$&0.8&0.35\\
$10^{-5}$&1.4&0.35\\
$10^{-5}$&2.0&0.35\\
0.0001&0.8&0.35\\
0.0001&1.4&0.35\\
0.0001&2.0&0.35\\
0.001&0.8&0.34\\
0.001&1.4&0.35\\
0.001&2.0&0.35\\
 \hline
 \end{tabular}
      \hspace{.5cm}%
  }
  \caption{Minimal values of $\lambda_{Hs}$ needed to stabilise the Higgs potential in the CSI ESM $\oplus$ singlet models
  with $\lambda_s=0.1$ and $\lambda_{\phi s}= 0.01$.
Left Table:  U(1)$_{\bf \sst  B-L}$. Right Table: SU(2)$_{\sst \mathrm{CW}}$.} \label{U1BLSStab}
\end{table}

%%%%%%%%%%%%%%%%%%%%%%%%%%%%%%%%%%%%%%%%%%%%%%%%%%%%
\subsection{CSI \texorpdfstring{SU(2)$_{\bf \sst \mathrm{CW}}\,\times$ \!SM}{SU(2)CWxSM}}
\label{sec:4.4}
%%%%%%%%%%%%%%%%%%%%%%%%%%%%%%%%%%%%%%%%%%%%%%%%%%%%

\begin{figure}[t!]
\centering
\includegraphics[width=.75\columnwidth]{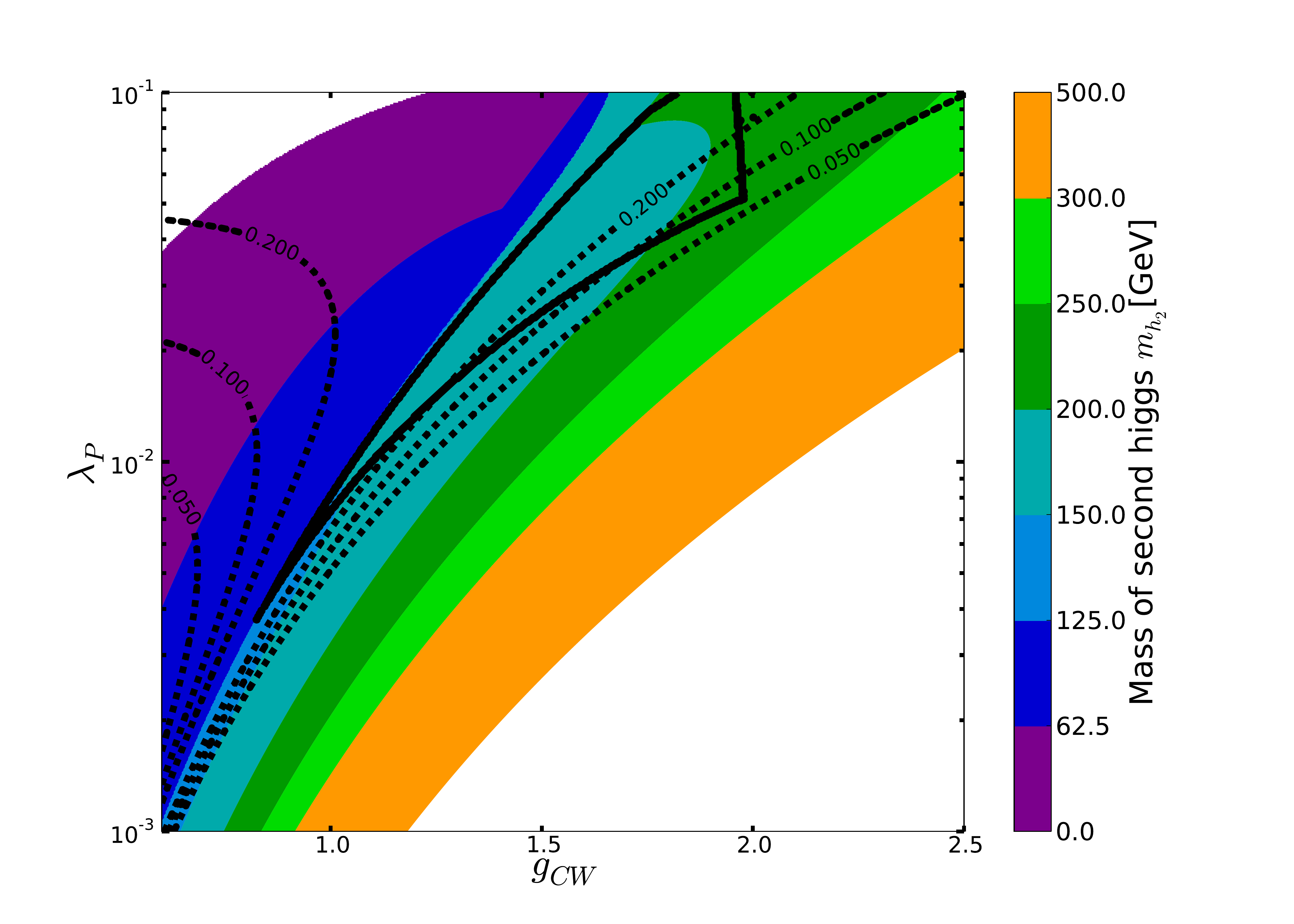}
\caption{Parameter space of the
 SU(2)$_{\sst \mathrm{CW}}\,\times$ \!SM theory showing the region
where the Higgs potential is stabilised and the 
 $\sin^2 \theta$ contours.
The legend is the same as in Figure~\ref{U1Stab}.}
\label{SU2Stab}
\end{figure}

Solving RG equations in the non-Abelian CW theory coupled to the SM, gives the Higgs stability region
shown in fig.~\ref{SU2Stab} together with the $\sin^2 \theta$ exclusion contours. The stability wedge is now
shifted to larger values of $g_{\sst \mathrm{CW}}$ as $\phi$ has an equivalent charge of $1/2$. 
From fig.~\ref{SU2Stab} we conclude that 
the combination of the Higgs potential stabilisation and the LHC limits on the heavier Higgs leaves a small
corner of the parameter space available 
in the minimal  SU(2)$_{\sst \mathrm{CW}}\times$ SM theory.

%%%%%%%%%%%%%%%%%%%%%%%%%%%%%%%%%%%%%%%%%%%%%%%%%%%%
\subsection{CSI \texorpdfstring{SU(2)$_{\bf \sst \mathrm{CW}}\,\times$ \!SM~$\oplus$}{SU(2)CWxSMplus} singlet}
\label{sec:4.5}
%%%%%%%%%%%%%%%%%%%%%%%%%%%%%%%%%%%%%%%%%%%%%%%%%%%%

The Higgs potential in the SU(2)$_{\sst \mathrm{CW}}\times$ SM can be stabilised on the entire 2d plane $(\lambda_{\rm P}, g_{\sst \mathrm{CW}})$ 
by extending the model with a vev-less singlet $s(x)$ portally coupled to the Higgs, as in~eq.~\eqref{bhs}.
The  table on the right in Table \ref{U1BLSStab} shows the critical value of $\lambda_{Hs}$ for this stabilisation mechanism to work
in the CSI SU(2)$_{\bf \sst \mathrm{CW}}\times \mathrm{SM}$ $\oplus$ singlet model.

Before we conclude this section we would like to make a comment.
We have shown that the minimal Higgs portal models without an additional scalar are largely
ruled out by the combination of Higgs (in)stability and the LHC constraints (except for a small region of the
parameter space still available in the non-Abelian model). At the same time we showed that if these models include an additional scalar field with a portal coupling
$\lambda_{Hs} \sim 0.35$, the Higgs stability restrictions are completely lifted and the models are completely viable.

The question arises if this conclusion would also apply to models without an additional scalar, but instead with the
Higgs-CW portal coupling being relatively large, $\lambda_{\rm P} \sim 0.3$, 
so that $\beta_{\lambda_H}$ would instead receive a positive contribution from 
$2 \lambda_{\rm P}^2$. This approach would not work for the following reason.
In order not to get a large mixing angle $\sin^2\theta>0.1$ in this case we require that the second scalar 
is  quite heavy, $m_{h_2}>300$ GeV. This in turn requires a large CW gauge coupling of $g_{\sst \mathrm{CW}} \approx 3.5$.
 Such a large gauge coupling leads to a large value for $\lambda_\phi$ at the scale of $\langle\phi\rangle$. 
 $\lambda_\phi$ therefore develops a Landau pole already at low scales.

%%%%%%%%%%%%%%%%%%%%%%%%%%%%%%%%%%%%%%%%%%%%%%%%%%%%
%%%%%%%%%%%%%%%%%%%%%%%%%%%%%%%%%%%%%%%%%%%%%%%%%%%%
\section{Dark Matter Physics: relic abundance and constraints}
\label{sec:DM}
%%%%%%%%%%%%%%%%%%%%%%%%%%%%%%%%%%%%%%%%%%%%%%%%%%%%
%%%%%%%%%%%%%%%%%%%%%%%%%%%%%%%%%%%%%%%%%%%%%%%%%%%%

Having demonstrated that the Higgs sector can be stabilised and that it is in agreement with all current observations, we now show that this framework can accommodate the observed dark matter content in the Universe. In the scenarios that we have studied, there are two potential dark matter candidates. The first candidate is the vector dark matter 
\cite{Hambye:2008bq,Hambye:2009fg,Arina:2009uq}
given by the triplet of
gauge bosons $Z'_i$ of the SU$(2)_{\sst \mathrm{CW}}$ sector and considered recently in \cite{Hambye:2013dgv,Carone:2013wla}.
These particles have the same mass $M_{Z'}$ and are stable because of an unbroken global $SO(3)$ `custodial symmetry', which also ensures that each component has the same relic abundance. The second candidate is the singlet scalar particle $s$ coupled to the Higgs through the Higgs portal.\footnote{Magnetic monopoles are also a possible third dark matter candidate~\cite{Evslin:2012fe}; in this work we ignore this possibility.}  This is a much studied dark matter candidate 
\cite{Silveira:1985rk,S2,S3,Mambrini:2011ik,Djouadi:2011aa,Low:2011kp,Cheung:2012xb,Cline:2013gha}
 that is stable because of a  $Z_2$ symmetry 
of the classically scale-invariant $\mathrm{SM}\times$G$_{\sst \mathrm{CW}}$ theory
with the real singlet \cite{Khoze:2013uia}.\footnote{The $s \to -s$ symmetry
of the potential eq.~\eqref{potentialcoupled3}
is an automatic consequence of scale-invariance and gauge invariance, 
which does not allow odd powers of $H$ and $\Phi$.} 

Having argued that the vector triplet and scalar particles are stable and therefore potential dark matter candidates, we must calculate the relic abundance in order to show that they can saturate, or form a component of the observed dark matter abundance, for which we take $\Omega_{\rm{DM}} h^2=0.1187\pm0.0017$, the value inferred from Planck+WP+HighL+BAO data~\cite{Ade:2013zuv}. Owing to the reasonable couplings to the Standard Model particles, the scalar and vector dark matter components
are in thermal equilibrium with the Standard Model degrees of freedom in the early Universe. Their abundance is therefore determined by the thermal freeze-out mechanism. To calculate it, we must solve the Boltzmann equation, which is~\cite{Gondolo:1990dk, D'Eramo:2010ep},
\begin{equation}
\frac{d n_i}{dt}+3Hn_i=-\langle\sigma_{ii}v\rangle\left(n_i^2-n_i^{\rm{eq}\,2} \right)-\sum_{j,k}\langle\sigma_{ijk}v\rangle\left(n_i n_j-\frac{n_k}{n_k^{\rm{eq}}}n_i^{\rm{eq}}n_j^{\rm{eq}}\right)\;,
\end{equation}
where $n_i$ is the number density of one component $\chi_i$ of the dark matter abundance, $\langle\sigma_{ii}v\rangle$ is the usual annihilation cross-section term for reactions of the form $\chi_i \chi_j\to X X$, where $X$ is a particle in equilibrium with the thermal bath,
 and $\langle\sigma_{ijk}v\rangle$ is the cross-section for the semi-annihilation reaction $\chi_i \chi_j\to \chi_k X$.

%%%%%%%%%%%%%%%%%%%%%%%%%%%%%%%%%%%%%%%%%%%%%%%%%%%%
\subsection{Vector dark matter}
\label{sec:vDM}
%%%%%%%%%%%%%%%%%%%%%%%%%%%%%%%%%%%%%%%%%%%%%%%%%%%%

\begin{figure}[t!]
\centering
\includegraphics[width=0.99 \columnwidth]{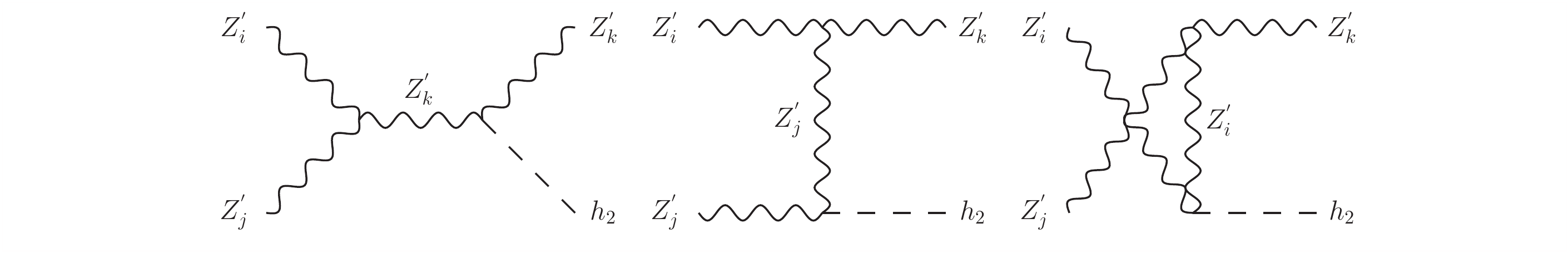}
\includegraphics[width=0.99 \columnwidth]{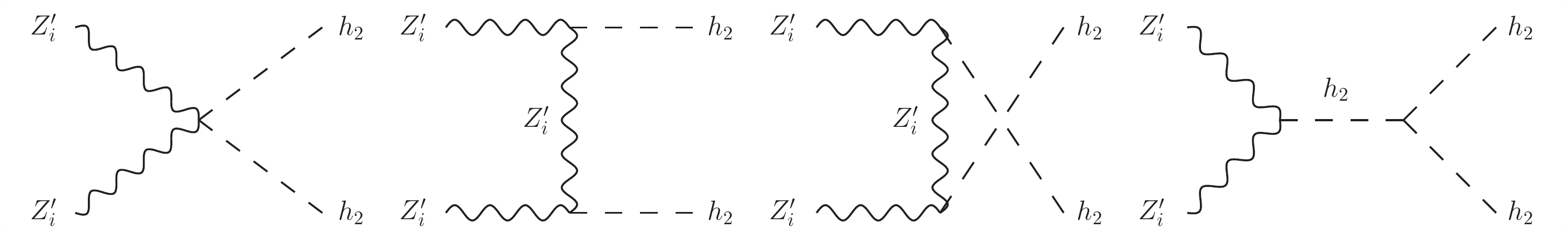}
\caption{
The upper three diagrams show the process $Z'_i Z'_j\to Z'_k h_2$, which is the dominant contribution to the semi-annihilation cross-section. The process $Z'_i Z'_j\to Z'_k h_1$ also occurs but is suppressed by $\tan^2\theta$.  The lower four diagrams show the processes that dominate the annihilation of $Z'_i Z'_i$. Other diagrams are suppressed by at least one power of $\sin\theta$ or $\lambda_{\rm{P}}$.}
\label{fig:semiann}
\end{figure}

We first consider the case of vector dark matter only, which is similar to Hambye's model~\cite{Hambye:2008bq} except that here there are no explicit $\mu$ terms. This model is interesting as it was the first example of a model containing both annihilation and semi-annihilation processes, as shown in fig.~\ref{fig:semiann}. 

The annihilation cross-section is dominated by the lower four diagrams of fig.~\ref{fig:semiann}, which contribute to the process $Z^{'}_i Z^{'}_i \to h_2 h_2$. The leading order terms contributing to the non-relativistic (s-wave) cross-section from these diagrams are 
\begin{equation}
\label{eq:xsecij}
\langle\sigma_{ii} v\rangle=\frac{11g_{\sst \mathrm{CW}}^4-60 g_{\sst \mathrm{CW}}^2 \lambda_{\phi}+108 \lambda_{\phi}^2}{2304\pi}\,\frac{ \cos^4\theta}{ M^2_{Z'}}+\mathcal{O}\left(\frac{m^2_{h_2}}{M^2_{Z'}},\sin\theta,\lambda_{\rm{P}}\right)\;.
\end{equation}
In our numerical work, we include all sub-leading terms in this cross-section as well as including the contributions from $Z'_i Z'_i\to h_1 h_1$, $Z'_i Z'_i\to\bar{f}f$, $Z'_i Z'_i\to W^{+}W^{-}$ and $Z'_i Z'_i\to Z^{0}Z^{0}$, all of which are suppressed by at least one power of $\sin\theta$ or $\lambda_{\rm{P}}.$

The diagrams that contribute to the semi-annihilation process are shown by the upper three diagrams in fig.~\ref{fig:semiann}. In the non-relativistic limit, the (s-wave) cross-section for $Z'_i Z'_j\to Z'_k h_2$ is
\begin{equation}
\label{eg:xsecijk}
\langle\sigma_{ijk} v\rangle=\frac{3 g_{\sst \mathrm{CW}}^4}{128 \pi}\frac{\cos^2\theta}{ M_{Z'}^2}\left(1-\frac{m_{h_2}^2}{3 M_{Z'}^2}\right)^{-2}\left(1-\frac{10 m_{h_2}^2}{9 M_{Z'}^2} +\frac{m_{h_2}^4}{9 M_{Z'}^4}\right)^{3/2}\;.
\end{equation}
There is also a subdominant process $Z'_i Z'_j\to Z'_k h_1$ whose cross-section is obtained from eq.~\eqref{eg:xsecijk} by substituting $m_{h_2}\to m_{h_1}$ and $\cos\theta\to\sin\theta$. For completeness, we include this in our numerical work. Comparing eqs.~\eqref{eq:xsecij} and \eqref{eg:xsecijk}, we observe that $\langle\sigma_{ijk} v\rangle\sim5\langle\sigma_{ij} v\rangle$ so the semi-annihilation processes dominate.

The global custodial symmetry ensures that the vector triplet is degenerate in mass and each $Z'_i$ contributes one-third to the relic abundance. That is the total abundance $n_{Z'}$ is related to the individual components by $n_{Z'}=3 n_{Z'_1}=3 n_{Z'_2}=3 n_{Z'_3}.$
It should also be clear that $\langle\sigma_{11}v\rangle=\langle\sigma_{22}v\rangle=\langle\sigma_{33}v\rangle:=\langle\sigma v\rangle_{\rm{ann}}$ and $\langle\sigma_{123}v\rangle=\langle\sigma_{132}v\rangle=\langle\sigma_{213}v\rangle=\langle\sigma_{231}v\rangle=\langle\sigma_{312}v\rangle=\langle\sigma_{321}v\rangle:=\langle\sigma v\rangle_{\rm{semi-ann}}$. Therefore, the Boltzmann equation for the total abundance is
\begin{equation}
\label{eq:Boltzvect}
\frac{d n_{Z'}}{dt}+3H n_{Z'}=-\frac{\langle\sigma v\rangle_{\rm{ann}}}{3}\left(n_{Z'}^2-n_{Z'}^{\rm{eq}\,2} \right)-\frac{2\langle\sigma v\rangle_{\rm{semi-ann}}}{3}n_{Z'}\left(n_{Z'} - n_{Z'}^{\rm{eq}}\right)\;.
\end{equation}
We solve this equation numerically by the method outlined in~\cite{Belanger:2012vp}.

\begin{figure}[t!]
\centering
\includegraphics[width=0.49\columnwidth]{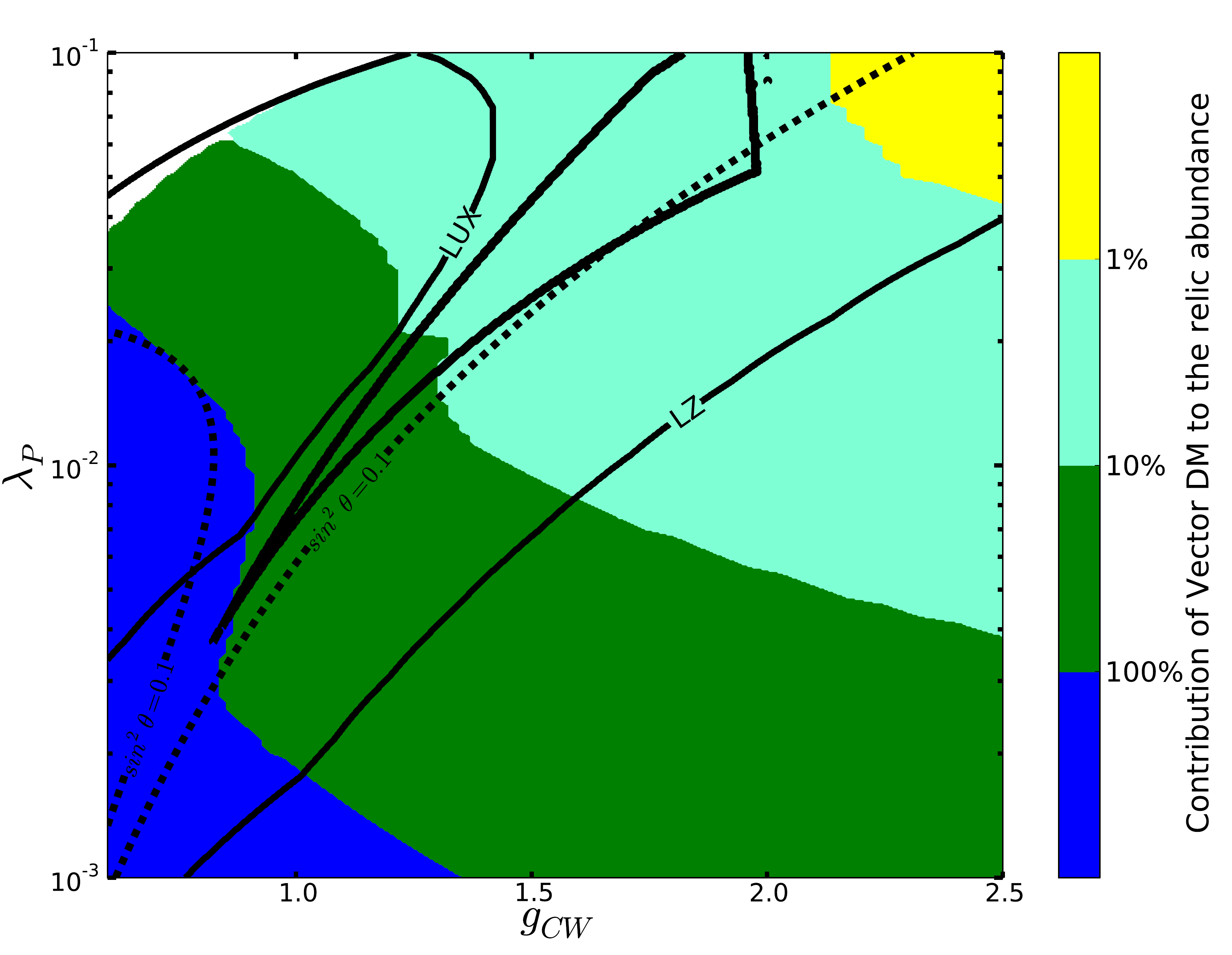}\hspace{2mm}\includegraphics[width=0.49 \columnwidth]{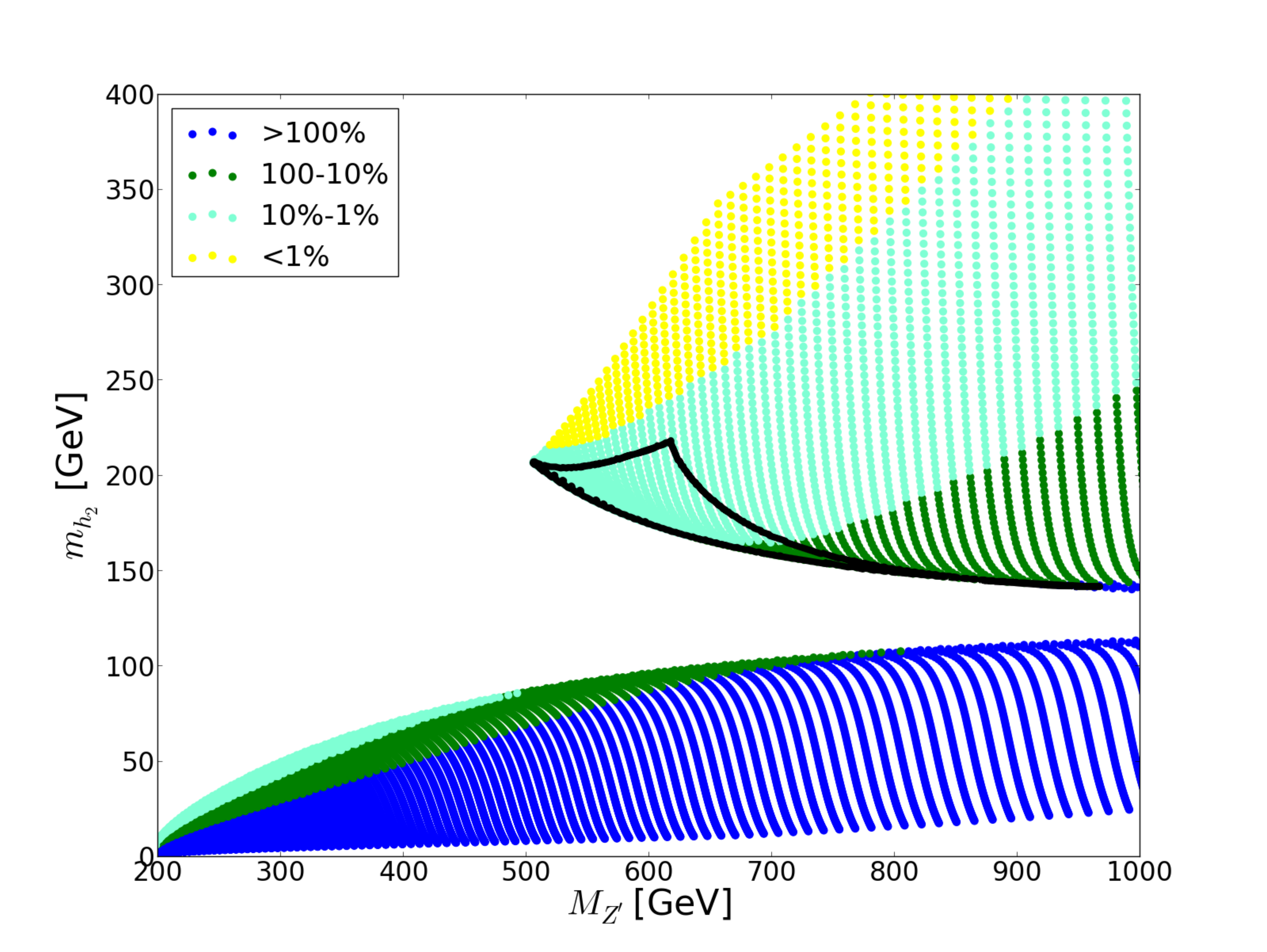}
\caption{
The coloured contours and the wedge-shaped regions in black in both panels indicate when the vector triplet forms more less than 100\%, 10\% and 1\% of the observed dark matter abundance, and the parameter values where the Higgs potential is stabilised respectively. Also shown in the left panel are the LUX and projected LZ limits (the region above these lines is excluded), which account for the fact that the dark matter is a subcomponent of the total density in much of the parameter space, and the limit $\sin^2\theta=0.1$. The right panel shows that the vector mass should lie between 500~GeV and 1~TeV to improve Higgs stability.}
\label{fig:vector_relic}
\end{figure}

The coloured regions in the left and right panels of fig.~\ref{fig:vector_relic} show the total relic abundance of the vector triplet as a fraction of the observed abundance. For instance, in the lower left  (blue) part of the left panel, the abundance exceeds the observed value and is therefore excluded. The thick black wedge indicates the region where the Higgs potential is stabilised up to the Planck scale (as in fig.~\ref{SU2Stab}). We see that for most of wedge, the vector triplet contributes between 1\% and 100\% of the total dark matter abundance. 
However, when we combine this with the LHC constraint on $\sin^2 \theta$, we see from fig.~\ref{fig:vector_relic} that the vector
dark matter component contributes less than 10\% to the total relic abundance, and we need to add another dark matter component.
The right panel in fig.~\ref{fig:vector_relic} shows the dark matter fraction as a function of $M_{Z'}$ and $m_{h_2}$. To lie within the
Higgs vacuum stability wedge, we see that the $M_{Z'}$ lies between 500~GeV and 1000~GeV.

Also shown on the left panel are the direct detection current constraints from LUX~\cite{Akerib:2013tjd} and the projected limits from LZ~\cite{Cushman:2013zza}. At a direct detection experiment, a vector $Z'_i$ can elastically scatter with a nucleon $N$ via exchange of $h_1$ or $h_2$. The resulting spin-independent scattering cross-section for this to occur is
\begin{equation}
\sigma_{N}^{\rm{SI}}=\frac{g_{\sst \mathrm{CW}}^2 \sin^2 2\theta }{16 \pi}\frac{f_N^2 m_{N}^2 \mu_{\rm{red}}^2}{v^2}\left(\frac{1}{m_{h_2}^2}-\frac{1}{m_{h_1}^2} \right)^2\;,
\end{equation}
where $f_N:=\bra{N}\sum_q m_q \bar{q}q \ket{N}/m_N \approx 0.295$ is the Higgs-nucleon coupling~\cite{Cheng:2012qr}, $m_N$ is the nucleon mass and $\mu_{\rm{red}}$ is the vector-nucleon reduced mass. When setting a limit from the experimental data, we account for the fact that the the vector triplet forms a subcomponent of the total dark matter density over much of the parameter space of interest. We make a scaling ansatz that the fraction of the local dark matter density $\rho_{Z'}/\rho_{\rm{DM}}$ is the same as the fraction of the dark matter relic abundance $\Omega_{Z'}/\Omega_{\rm{DM}}$. The limits from LUX and LZ after taking into account this scaling are shown in fig.~\ref{fig:vector_relic} by the lines with the appropriate label. In the left panel, the region above and to the left of the lines are excluded. 
We have also checked that the LUX exclusion limit, when applied to the right panel, excludes the entire lower island. 
Therefore, while the current LUX limits do not constrain the region where the Higgs potential is stabilised, the projected LZ limit excludes all of this region.

%%%%%%%%%%%%%%%%%%%%%%%%%%%%%%%%%%%%%%%%%%%%%%%%%%%%
\subsection{Singlet scalar dark matter}
\label{sec:sDM}
%%%%%%%%%%%%%%%%%%%%%%%%%%%%%%%%%%%%%%%%%%%%%%%%%%%%

\begin{figure}[t!]
\centering
\includegraphics[width=1.0 \columnwidth]{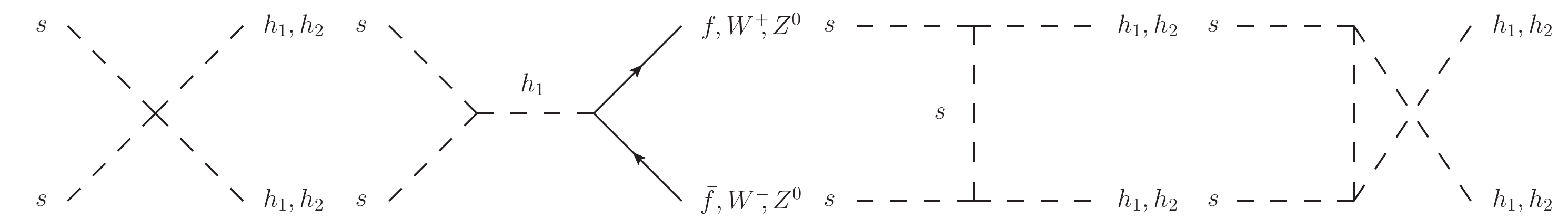}
\caption{
The leading contributions to the scalar annihilation cross-section $\langle\sigma v\rangle_{\rm{s,ann}}$. Other diagrams are suppressed by at least one power of $\sin\theta$.}
\label{fig:scalann}
\end{figure}

We have previously motivated the introduction of a real singlet scalar field to allow the Higgs potential to be stabilised over a much larger range of the parameter space. Providing a candidate to saturate the observed dark matter abundance provides a second motivation. The two examples of CSI ESM with a U$(1)$ Coleman-Weinberg sector that we have considered 
in sections {\bf \ref{sec:4.1}} and {\bf \ref{sec:4.2}},
do not have a dark matter candidate. This is because the U(1)$_{\sst \mathrm{CW}}$ gauge boson is unstable, owing to its kinetic mixing with
hypercharge, and the only scalar field present, $\phi_{\sst \mathrm{CW}},$ mixes with the SM Higgs. 
The SU(2)$_{\sst \mathrm{CW}}$ sector does have a stable component in the form of the $Z'_i$ triplet, but we have seen 
cf.~left panel in fig.~\ref{fig:vector_relic})
that after LHC constraints have been taken into account, the vector triplet forms only a sub-component of the total dark matter abundance in the region where the Higgs potential is stabilised. Therefore, in the case of an SU$(2)$ extended Standard Model, an additional dark matter component is also required. 

Having motivated the singlet scalar as a dark matter candidate, we first study the case where the singlet forms all of the dark matter (as required in the U(1) case) before turning to the case where it forms a sub-component (as required in the SU(2) case).

\begin{figure}[t!]
\centering
\includegraphics[width=0.75\columnwidth]{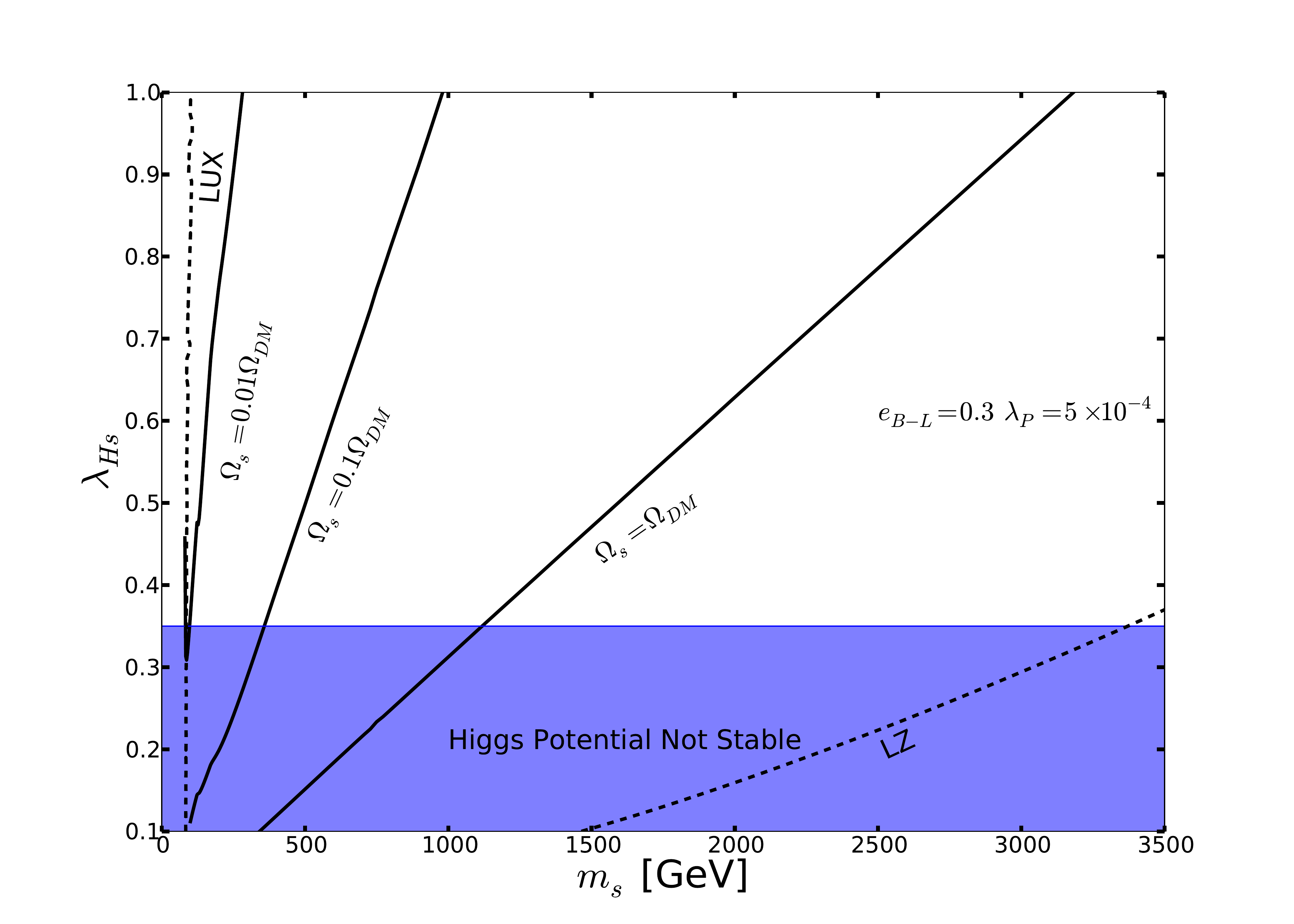}
\caption{Scalar dark matter $(m_s,\lambda_{Hs})$ plane in the CSI U(1)$_{\bf \sst B-L}\,\times$ \!SM $\oplus$ singlet model.
The solid lines show the fraction of of the total DM density
the scalar singlet makes up. The dotted lines show the direct detection
constraints from LUX and the project limits from LZ. In the shaded
region the extra singlet does not stabilise the Higgs potential.}
\label{fig:scal_omega}
\end{figure}

In the CSI U(1)${_{\bf \sst  B-L}}\times$ SM $\oplus$ singlet model, the ATLAS and CMS limit that \mbox{$M_{Z'}\gtrsim3$~TeV} implies that $\lambda_{\rm{P}}$, and therefore $\sin\theta$, is small. As a result, the diagrams that dominantly contribute to the total annihilation cross-section $\langle\sigma v\rangle_{\rm{s,ann}}$ are those shown in fig.~\ref{fig:scalann}. The $Z_2$ symmetry of this theory ensures that all semi-annihilation processes vanish, so that the Boltzmann equation describing the evolution of the scalar number density $n_s$ is the usual one:
\begin{equation}
\label{eq:Boltzscal}
\frac{d n_{s}}{dt}+3H n_{s}=-\langle\sigma v\rangle_{\rm{s,ann}}\left(n_{s}^2-n_{s}^{\rm{eq}\,2} \right)\;.
\end{equation}
The main parameters of our singlet dark models are the scalar dark matter mass, $m_s,$ and its coupling, $\lambda_{Hs},$ to the Higgs field.
We solve the Boltzmann equation numerically and the results are displayed in fig.~\ref{fig:scal_omega}
on the $(m_s,\lambda_{Hs})$ plane. 
In this figure, we have initially fixed $e_{\bf \sst B-L}=0.3$ and $\lambda_{\rm{P}}=5\times10^{-4}$ resulting in a mixing angle $\theta\approx5\times10^{-3}$ and mass $M_{Z'}=3.6$~TeV. When $e_{\bf \sst B-L}$ and $\lambda_{\rm{P}}$ are chosen so that $M_{Z'}$ lies above the bounds from direct searches by ATLAS and CMS, we have found that the positions of the lines are not sensitive to the values of $e_{\bf \sst B-L}$ and $\lambda_{\rm{P}}$. The coupling constant $\lambda_{\phi s}$ can be traded in for $m_s^2$ cf.~eq.~\eqref{ms2}) so that the only remaining free parameters are $m_s$ and $\lambda_{hs}$ (the quadratic coupling $\lambda_s$ plays no role in the Born-level freeze-out calculation). For each value of $m_s$, the value of $\lambda_{Hs}$ that gives 100\%, 10\% or 1\% of the observed dark matter density $\Omega_{\rm{DM}}$ is shown in fig.~\ref{fig:scal_omega}. The region below $\lambda_{Hs}\sim0.34$ is excluded because for these values of $\lambda_{Hs}$, the real scalar does not help to stabilise the Higgs potential cf.~Table~\ref{U1BLSStab}. We also impose that $\lambda_{Hs}\lesssim1$ in order that $\lambda_{Hs}$ does not develop a Landau pole before the Planck scale. In order that the singlet scalar saturates the observed dark matter density, we find that its mass should lie in the range between 1~TeV and 3.2~TeV. In this range, the annihilation channel $ss\to Z' Z'$ is not allowed kinematically, justifying its exclusion from the diagrams in fig.~\ref{fig:scalann}.

Finally, we also show the current direct detection constraints from LUX and the projected limits from LZ. The scalar can scatter at a direct detection experiment through t-channel exchange of $h_1$ and $h_2$ and the resulting spin-independent scattering cross-section to scatter off a nucleon $N$ is
\begin{equation}
\sigma_{N}^{\rm{SI}}=\frac{\lambda_{Hs}^2 \cos^4\theta }{4 \pi}\frac{f_N^2 m_{N}^2 \mu_{\rm{red}}^2}{m_s^2 m_{h_1}^4}\left[1-\tan\theta \left(\frac{\lambda_{\phi s}}{\lambda_{Hs}}-\frac{m_{h_1}^2}{m_{h_2}^2}\left(\frac{\lambda_{\phi s}}{\lambda_{Hs}}+\tan\theta\right) \right) \right]^2\;.
\end{equation}
As in the case of the vector triplet, we account for the fact that the scalar makes up a sub-component of the dark matter in much of the parameter space. While the current LUX limit constrains low values of $m_s$ where the scalar density $\Omega_s$ is very low, the projected LZ limits should constrain the full parameter space of interest.

\begin{figure}[t!]
\centering
\includegraphics[width=0.49\columnwidth]{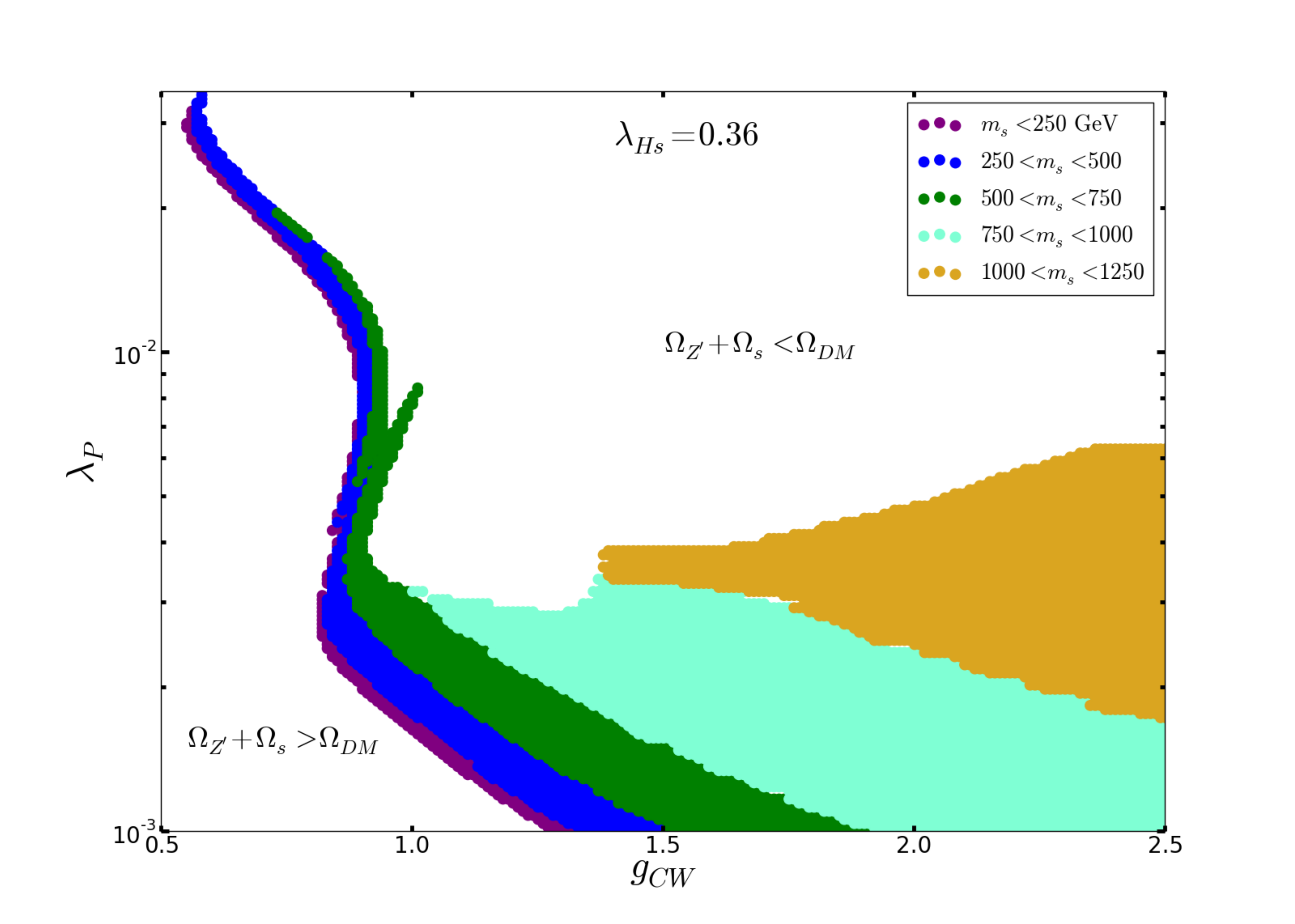}\hspace{2mm}\includegraphics[width=0.49 \columnwidth]{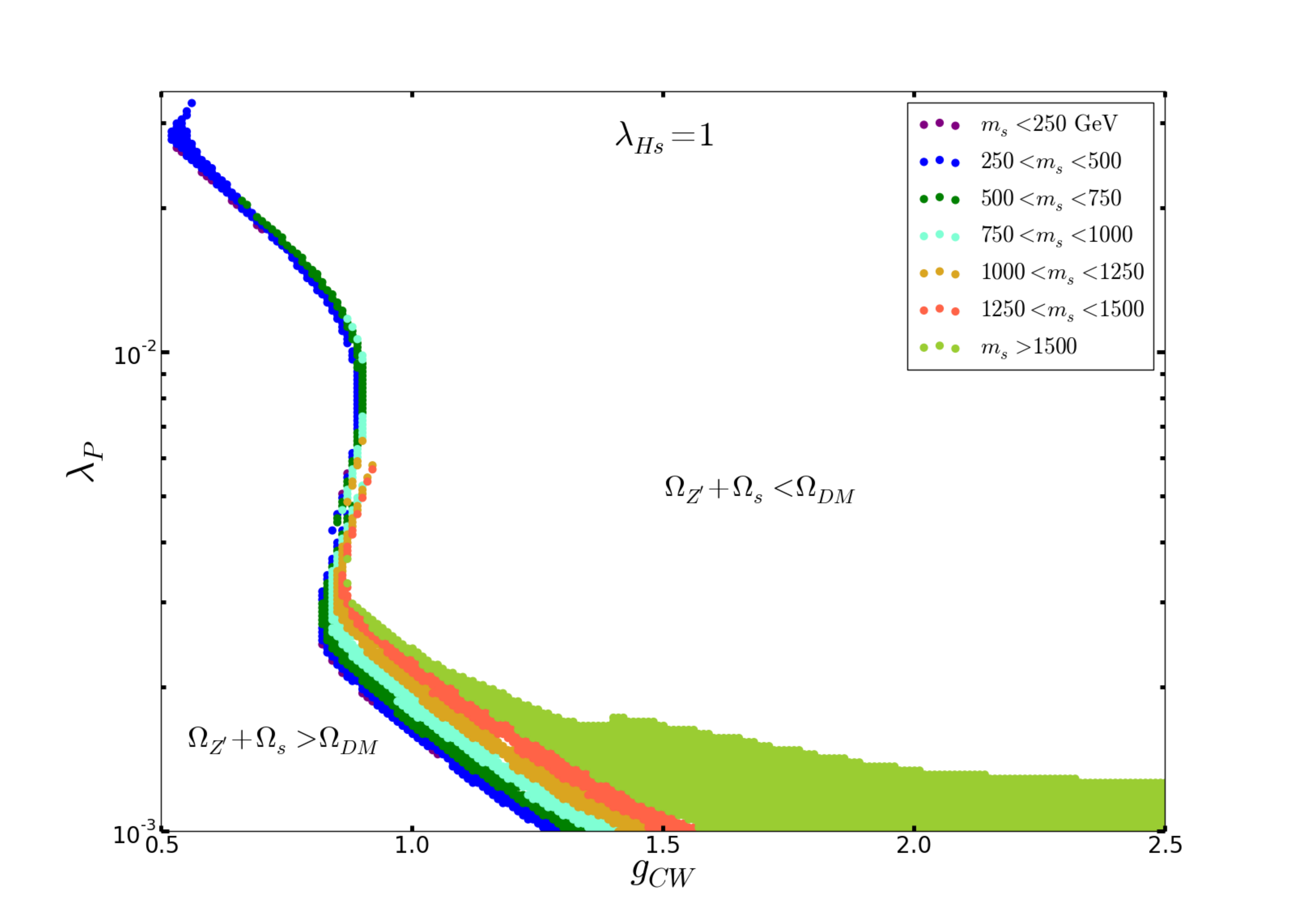}
\caption{The plots show the available parameter space when the scalar
and vector dark matter together makes up the total dark matter density in the of the CSI SU(2)$_{\sst \mathrm{CW}}\,\times$ \!SM $\oplus$ singlet model.
The colour-coded regions show the scalar dark matter mass in GeV. In the white
regions the combined density is either larger or smaller than the
observed dark matter density. On the left we fixed $\lambda_{Hs}=0.36$,
and the right panel has $\lambda_{Hs}=1$.}
\label{fig:mix_relic}
\end{figure}

%%%%%%%%%%%%%%%%%%%%%%%%%%%%%%%%%%%%%%%%%%%%%%%%%%%%
\subsection{Scalar and vector dark matter}
\label{sec:svDM}
%%%%%%%%%%%%%%%%%%%%%%%%%%%%%%%%%%%%%%%%%%%%%%%%%%%%

Finally, we consider the CSI SU(2)$_{\sst \mathrm{CW}}\times$~SM~$\oplus$~singlet model in which the dark matter is comprised of both the singlet scalar and vector triplet. In this case we solve the Boltzmann equations~\eqref{eq:Boltzvect} and~\eqref{eq:Boltzscal} as before, but we now include the annihilation process $ss\to Z'_i Z'_i$ or the reverse process, depending on which is kinematically allowed.

Figure~\ref{fig:mix_relic} shows
the results in the $(g_{\sst \mathrm{CW}},\lambda_{\rm P})$ plane for $\lambda_{Hs}=0.36$ and $\lambda_{Hs}=1.0$ in the left and 
right panels respectively. The coloured contours indicate the values of $m_s$ that results in the total density of vector and scalar saturating the observed value i.e.~$\Omega_{Z'}+\Omega_s=\Omega_{\rm{DM}}$. There is a limited portion of the parameter space in which the vector and scalar make up all of the dark matter and this region is smaller in the case where $\lambda_{Hs}$ is bigger. These results can be understood with reference to 
figs.~\ref{fig:vector_relic} and~\ref{fig:scal_omega}. From Figure~\ref{fig:vector_relic}, we observe that in the upper right corner of the left panel, the vector density is very small, so that the scalar should make up most of the density. From the right panel, we also see that  in this region, $M_{Z'}\lesssim1$~TeV, which because $g\approx2$, implies that $\langle \phi \rangle\lesssim1$~TeV. Now, from fig.~\ref{fig:scal_omega}, we see that for $\lambda_{Hs}=0.36$, we require $m_s\approx1$~TeV in order that $\Omega_s\approx\Omega_{\rm{DM}}$. However, given that $m_s^2\approx\lambda_{\phi s}| \langle \phi \rangle|^2/\sqrt{2}$ (cf.~eq.~\eqref{ms2}), we see that we can not achieve $m_s\approx1$~TeV unless $\lambda_{\phi s}\gtrsim1$, in which case, it develops a Landau Pole before the Planck scale. Figure~\ref{fig:scal_omega} also allows us to see why the parameter space is smaller for larger $\lambda_{Hs}$. This is because the value of $m_s$ that is required to obtain $\Omega_s\approx\Omega_{\rm{DM}}$ is larger for larger $\lambda_{Hs}$ and this is more difficult to do, again because of the perturbativity restriction on $\lambda_{\phi s}$.

\begin{figure}[t!]
\centering
\includegraphics[width=0.7\columnwidth]{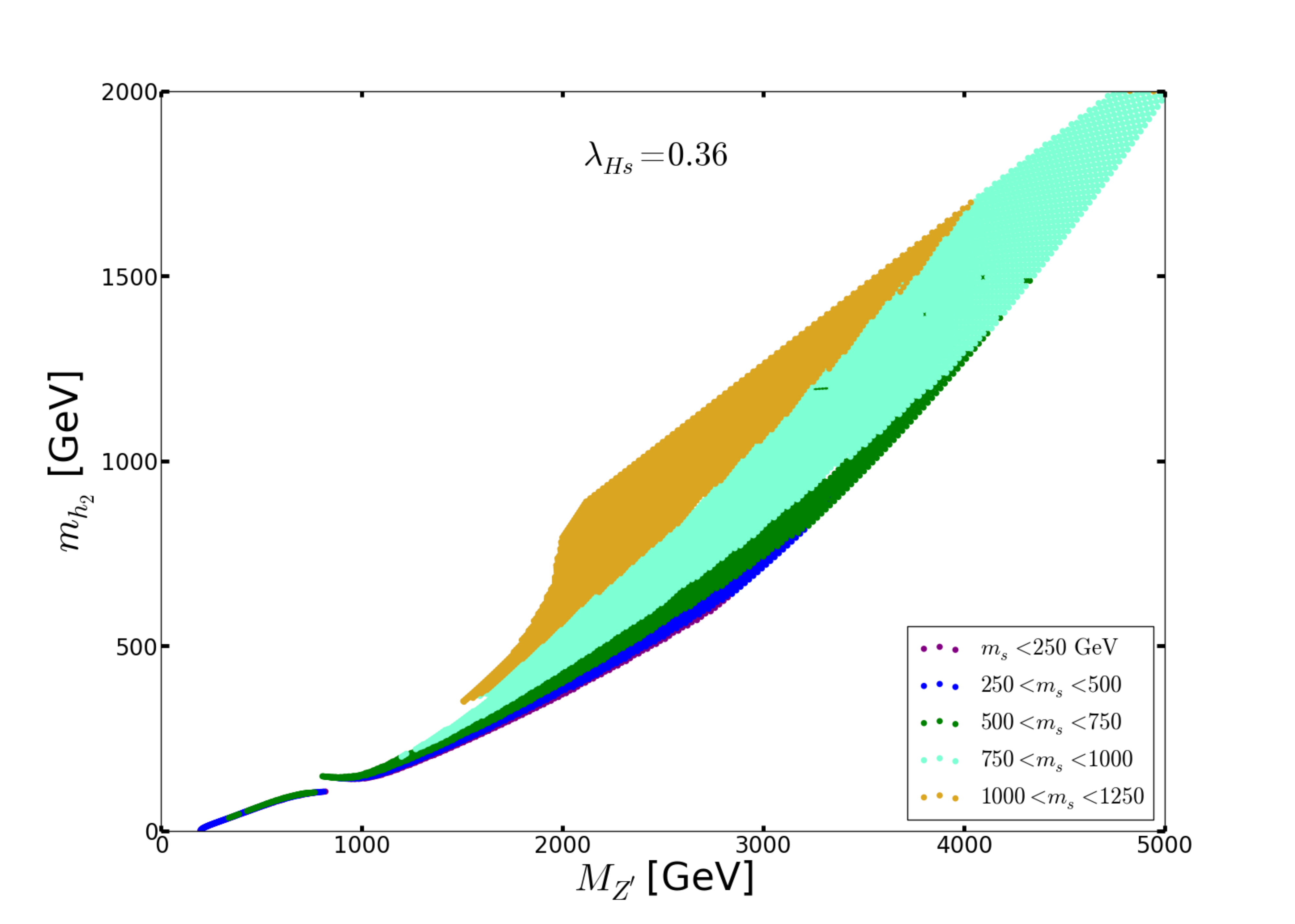}
\caption{The region on the mass plane $(M_{Z'},m_{h_2})$ where the combined density of the
scalar and vector dark matter equals the observed dark matter density.
The colours show the scalar dark matter mass in GeV and in the white
regions the combined density is either larger or smaller than the
observed dark matter density. Here we have fixed $\lambda_{Hs}=0.36$.
}
\label{fig:m_mix_relic}
\end{figure}

Figure~\ref{fig:m_mix_relic} shows the vector and Coleman-Weinberg scalar mass and contours of the scalar mass in which the total density is saturated. This plot has $\lambda_{Hs}=0.36$. We see that both the vector and scalar are required to be around the TeV scale.

%%%%%%%%%%%%%%%%%%%%%%%%%%%%%%%%%%%%%%%%%%%%%%%%%%%%
%%%%%%%%%%%%%%%%%%%%%%%%%%%%%%%%%%%%%%%%%%%%%%%%%%%%
\section{Conclusions}
\label{sec:concl}
%%%%%%%%%%%%%%%%%%%%%%%%%%%%%%%%%%%%%%%%%%%%%%%%%%%%
%%%%%%%%%%%%%%%%%%%%%%%%%%%%%%%%%%%%%%%%%%%%%%%%%%%%

The classically scale-invariant extensions of the Standard Model constitute a highly predictive and minimal 
model building framework. In this CSI ESM set-up, all mass scales have to be generated dynamically and should
therefore have a common origin. These models have to address all sub-Planckian shortcomings of the Standard Model.
In this paper we have analysed the CSI ESM theories from the perspective of solving the instability problem of the
SM Higgs potential and at the same time providing viable dark matter candidates.

In simple CSI models with Abelian hidden sectors, we identified regions of parameter space where the SM Higgs potential is stabilised 
all the way up to the Planck scale. These are the wedge-shaped regions in figs.~\ref{U1Stab} and \ref{U1BLStab}.
When combined with LHC constraints on heavier Higgs bosons we found that these regions did not survive (see 
dotted lines in figs.~\ref{U1Stab} and \ref{U1BLStab}).

In the case of a non-Abelian SU(2) hidden sector in fig.~\ref{SU2Stab} a small part of the parameter space with the stable Higgs potential 
is compatible with the LHC constraints.

We then argued that by adding a real scalar singlet with a portal coupling to the Higgs $\lambda_{Hs} \gtrsim 0.35,$ 
all of our CSI ESM models have a stable Higgs potential and are consistent with the LHC exclusion limits on extended Higgs
sectors.

For Abelian models the singlet of mass $m_s$ is the only dark matter candidate, and fig.~\ref{fig:scal_omega}
shows the available parameter space 
on the $(m_s, \lambda_{Hs})$ plane. If this singlet contributes 100\% of the total observed dark matter density, its mass lies between 1~TeV and 3~TeV.
The LUX direct detection limits do not yet constrain the model, however the projected reach of LZ would cover all of the
viable parameter space.

In non-Abelian models we have two components of dark matter -- the singlet and the hidden sector SU(2) gauge bosons, $Z'_i$.
Without the singlet, the combination of Higgs stability and LHC constraints implies that vector dark matter contributes
less than 10\% of the observed relic density, as can be seen in fig.~\ref{fig:vector_relic}.
Thus, to saturate the dark matter density and stabilise the Higgs potential we are required to have a singlet dark matter component.
Finally, we have investigated the phenomenology of two-component dark matter. The viable regions of parameter space are shown in
figs.~\ref{fig:mix_relic} and \ref{fig:m_mix_relic}. Typically, both components have mass close to 1~TeV.

We see that CSI ESM models are viable and predictive. They provide a non-trivial link
between the electroweak scale, including the Higgs vacuum stability, and the nature and origin of dark matter.
Furthermore, future dark matter direct detection and collider experiments will be able to explore a significant fraction of their parameter space.

%%%%%%%%%%%%%%%%%%%%%%%%%%%%% 
\acknowledgments
%%%%%%%%%%%%%%%%%%%%%%%%%%%%%

We would like to thank C.~Arina,  C.~Boehm, C.~Englert, J.~Jaeckel, M.~Spannowsky, A.~Strumia for useful discussions and correspondence.
This material is based upon work supported  by 
STFC through the IPPP grant ST/G000905/1. VVK acknowledges the support of the Wolfson Foundation and Royal Society
through a Wolfson Research Merit Award.  GR acknowledges the receipt of a Durham Doctoral Studentship.

\bibliographystyle{JHEP}

\end{document}